\UseRawInputEncoding 
\documentclass[twocolumn,tighten,times,psfig]{aastex62}
\usepackage{apjfonts}
\usepackage{mathrsfs}
\usepackage{amsmath}
\usepackage{url}
\usepackage[normalem]{ulem}
\usepackage{graphicx}

\usepackage{natbib}
\usepackage{color}
\bibpunct{(}{)}{;}{a}{}{,}

\newcommand\aproxgt{\mathrel{%
      \rlap{\raise 0.511ex \hbox{$>$}}{\lower 0.511ex \hbox{$\sim$}}}}
\newcommand\aproxlt{\mathrel{%
      \rlap{\raise 0.511ex \hbox{$<$}}{\lower 0.511ex \hbox{$\sim$}}}}

\newcommand{\ignore}[1]{}

\newcommand{\storm}{{\sc STORM}}
\newcommand{\swift}{{\it Swift}}
\newcommand{\hst}{{\it HST\,}}
\newcommand{\mdm}{{\sc MDM}}
\newcommand{\lamp}{{\sc LAMP}}

\newcommand{\ps}{{\sc PrepSpec}}
\newcommand{\memecho}{{\sc MEMEcho}}

\newcommand{\caramel}{{\sc caramel}}

\newcommand{\UBV}{{\it UBV}}

\newcommand{\ngc}{NGC\,5548}
\newcommand{\et}{et~al.}
\newcommand{\mbh}{\mbox{$M_{\rm BH}$}}
\newcommand{\msun}{\mbox{${\rm M}_\odot$}}
\newcommand{\kms}{\mbox{\rm km~s$^{-1}$}}

\newcommand{\lya}{\mbox{\rm Ly$\alpha$}}
\newcommand{\ha}{\mbox{\rm H$\alpha$}}
\newcommand{\hb}{\mbox{\rm H$\beta$}}
\newcommand{\civ}{\mbox{\rm C\,{\sc iv}}}
\newcommand{\siiv}{\mbox{\rm Si\,{\sc iv}}}
\newcommand{\heii}{\mbox{\rm He\,{\sc ii}}}
\newcommand{\nv}{\mbox{\rm N\,{\sc v}}}
\newcommand{\oiii}{\mbox{\rm [O\,{\sc iii}]}}
\newcommand{\caii}{\mbox{\rm Ca\,{\sc ii}}}

\newcommand{\dd}{\mbox{\rm d}}
\newcommand{\fracd}[2]{\frac{\displaystyle{#1}}{\displaystyle{#2}}}

\newcommand{\new}[1]{{  #1}}

\shorttitle{AGN \storm\ ~IX.\ Velocity--Delay Maps of \ngc}
\shortauthors{Horne, \et}

\begin{document}

\title{\Large Space Telescope and Optical Reverberation Mapping Project.\\
IX.\ Velocity--Delay Maps for Broad Emission Lines in \ngc}


\author[0000-0003-1728-0304]{Keith~Horne}
\affiliation{\ignore{SUPA}SUPA Physics and Astronomy, University of St. Andrews, Fife, KY16 9SS Scotland, UK}

\author[0000-0003-3242-7052]{G.~De~Rosa}
\affiliation{Space Telescope Science Institute, 3700 San Martin Drive, Baltimore, MD 21218, USA}

\author[0000-0001-6481-5397]{B.~M.~Peterson}
\affiliation{Space Telescope Science Institute, 3700 San Martin Drive, Baltimore, MD 21218, USA}
\affiliation{Department of Astronomy, The Ohio State University, 140 W 18th Ave, Columbus, OH 43210, USA}
\affiliation{Center for Cosmology and AstroParticle Physics, The Ohio State University, 191 West Woodruff Ave, Columbus, OH 43210, USA}


\author[0000-0002-3026-0562]{A.~J.~Barth}
\affiliation{\ignore{UCI}Department of Physics and Astronomy, 4129 Frederick Reines Hall, University of California, Irvine, CA 92697, USA}

\author[0000-0002-4814-5511]{J.~Ely}
\affiliation{Space Telescope Science Institute, 3700 San Martin Drive, Baltimore, MD 21218, USA}

\author[0000-0002-9113-7162]{M.~M.~Fausnaugh}
\affiliation{Department of Astronomy, The Ohio State University, 140 W 18th Ave, Columbus, OH 43210, USA}
\affiliation{\ignore{}Kavli Institute for Space and Astrophysics Research,  Massachusetts Institute of Technology,  77 Massachusetts Avenue, Cambridge, MA 02139-4307, USA}

\author[0000-0002-2180-8266]{G.~A.~Kriss}
\affiliation{Space Telescope Science Institute, 3700 San Martin Drive, Baltimore, MD 21218, USA}

\author{L.~Pei}
\affiliation{\ignore{UCI}Department of Physics and Astronomy, 4129 Frederick Reines Hall, University of California, Irvine, CA 92697, USA}


\author[0000-0002-2816-5398]{M.~C.~Bentz}
\affiliation{\ignore{Georgia}Department of Physics and Astronomy, Georgia State University, 25 Park Place, Suite 605, Atlanta, GA 30303, USA}

\author[0000-0002-8294-9281]{E.~M.~Cackett}
\affiliation{\ignore{Wayne}Department of Physics and Astronomy, Wayne State University, 666 W. Hancock St, Detroit, MI 48201, USA}

% ...

\begin{abstract}
\new{
In this contribution, we achieve the primary goal of the active galactic nucleus (AGN) \storm\ campaign
by recovering velocity--delay maps for the prominent broad emission lines (\lya, \civ, \heii, and \hb)
in the spectrum of \ngc.
These are the most detailed velocity--delay maps ever obtained for an AGN,
providing unprecedented information on the geometry, ionization structure, 
and kinematics of the broad-line region.
Virial envelopes enclosing the emission-line responses show
that the reverberating gas is bound to the black hole.
A stratified ionization structure is evident.
The \heii\ response inside 5--10~light days has a broad single-peaked velocity profile.
The \lya, \civ, and \hb\ responses extend from inside 2 to outside 20 light days,
with double peaks at $\pm2500$~\kms\ in the 10--20~light-day delay range.
An incomplete ellipse in the velocity--delay plane is evident in \hb.
We interpret the maps in terms of a Keplerian disk
with a well-defined outer rim at $R=20$~light days.
The far-side response is weaker than that from the near side.
The line-center delay $\tau=(R/c)(1-\sin{i})\approx5$~days
gives the inclination $i\approx45^\circ$.
The inferred black hole mass is $\mbh\approx7\times10^7$~\msun.
In addition to reverberations, the fit residuals confirm
that emission-line fluxes are depressed during the
``BLR Holiday'' identified in previous work.
Moreover, a helical ``Barber-Pole'' pattern, with stripes moving from red to blue
across the \civ\ and \lya\ line profiles, suggests
azimuthal structure rotating with a 2~yr period that may represent
precession or orbital motion of inner-disk structures
casting shadows on the emission-line region farther out.
}

\end{abstract}


\section{Introduction}

\label{sec:intro}

\typeout{Intro}

Active galactic nuclei (AGN) are understood to be
powered by accretion onto supermassive black holes
in the nuclei of their host galaxies.
On account of angular momentum,
the accreting gas forms a disk on scales of a few to a few hundred
gravitational radii, $R_{\rm g} = G\, \mbh/c^2$, where
$\mbh$ is the mass of the central black hole.
The accretion disk ionizes gas on scales of hundreds to
thousands $R_{\rm g}$, which reprocesses the ionizing
radiation into strong emission lines that are significantly
Doppler-broadened by their motion in the deep gravitational
potential of the black hole. However, the structure and
kinematics of the ``broad-line region'' (BLR) remain among the
long-standing unsolved problems in AGN astrophysics.

It is generally supposed that the BLR plays some role in
the inflow and outflow processes that are known to
occur on these spatial scales. There is evidence for disk structure
in some cases
\citep[e.g.,][]{Wills86,Eracleous94,Vestergaard00,Eracleous03,
Strateva03,Smith04,Jarvis06,Gezari07,Young07,Lewis10,
Storchi-Bergmann17} as well as evidence that gravity dominates
the dynamics of the BLR  \citep[e.g.,][]{Peterson04},
although radiation pressure may also contribute \citep{Marconi08,Netzer10}.
Perhaps the strongest evidence for a BLR with black hole dominated motions 
and a thick-disk geometry
is the GRAVITY Collaboration's spectroastrometry results showing the red and blue wings 
of the P$\alpha$ line  spatially offset in opposite directions perpendicular to the jet
in the nearest quasar, 3C~273 \citep{Sturm18}.

The reverberation mapping  (RM) technique \citep{Blandford82,Peterson93,Peterson14}
affords a means of highly constraining the BLR geometry and kinematics
by measurement of the time-delayed response of the line flux to changes in the continuum
flux as a function of Doppler velocity. The projection of the BLR velocity field and
structure into the observables of Doppler velocity and time delay 
yields a ``velocity--delay map.'' 
Velocity--delay maps provide
detailed information on the BLR geometry,
velocity field, and ionization structure, and
can be constructed by analyzing the
reverberating velocity profiles \citep{Horne04}. 
This requires sustained monitoring
of the reverberating spectrum with
high signal-to-noise ratio (S/N) and high cadence
to record the subtle changes in the line profiles.

\subsection{ The 2014 \storm\ Campaign on \ngc. }

To secure data suitable
for velocity--delay mapping,
\ngc\ was the focus of an intensive monitoring campaign in
2014, the AGN Space Telescope and
Optical Reverberation Mapping (AGN STORM) program.
Ultraviolet (UV) spectra were obtained almost daily for
6 months with the Cosmic Origins Spectrograph on the
{\em Hubble Space Telescope} (\hst), 
securing 171 UV spectra covering rest-frame wavelengths 1130--1720\,\AA,
including the prominent \lya\,$\lambda1216$
and \civ\,$\lambda1549$ emission lines
and the weaker \siiv\,$\lambda1397$ and \heii\,$\lambda1640$
emission lines \citep[][hereafter Paper~I]{DeRosa15}.
During the middle two-thirds of the campaign,
observations with the \swift\ satellite provided longer-wavelength UV,
0.3 to 10~keV X-ray, and optical (\UBV) continuum
measurements \citep[][hereafter Paper~II]{Edelson15}.
A major ground-based campaign secured
imaging photometry \citep[][hereafter Paper~III]{Fausnaugh16}
with sub-diurnal cadence, including the
$UBV$ and Sloan $ugriz$ bandpasses. 
Optical spectroscopic
observations \citep[][hereafter Paper~V]{Pei17} 
were also obtained, \new{with 147 spectra} covering the Balmer
line \hb\,$\lambda4861$ and \heii\,$\lambda4686$.

Anomalous behavior in the emission-line response,
known colloquially as the BLR Holiday, is discussed by 
\citet[][hereafter Paper~IV]{Goad16}.
\citet[][hereafter Paper~X]{Dehghanian19} presents photoionization modeling
using the absorption lines to diagnose how the ionizing spectral energy
distribution changed during the BLR~Holiday.
Detailed fitting of a reverberating disk model to the \hst, \swift,
and optical light curves was accomplished 
by \citet[][hereafter Paper~VI]{Starkey16}.
The X-ray observations are discussed by
\citet[][hereafter Paper~VII]{Mathur17}.
A comprehensive analysis modeling of
the variable emission and absorption 
features is presented
by \citet[] [hereafter Paper~VIII]{Kriss19}.
The present manuscript, presenting velocity--delay maps derived from the
spectral variations, is Paper~IX.

Analysis of the 
\storm\ datasets has provided several breakthroughs
and surprises that challenge our previous understanding
of AGN accretion flows.
One major breakthrough is the first clear measurement of
interband continuum lags (Papers II and III), which
can serve as a probe of the accretion disk temperature profile
\citep{Collier01,Cackett07}.
This tests a key prediction of the standard \citet{SS73} disk theory,
$T_{\rm eff}\propto \left(\mbh\,\dot{M}\right)^{1/4}r^{-3/4}$,
where $\dot{M}$ is the accretion rate.
The \storm\ results are somewhat surprising, as follows.

\begin{itemize}

\item From the continuum and broad-band photometric
light curves, cross-correlation analysis
(Papers II and III),
and detailed light curve modeling (Paper~VI),
continuum lags and thus the disk size
are larger than expected, by a factor of $\sim3$.
Similarly, over-large disks are inferred from
microlensing effects in lensed quasar light curves
\citep{Poindexter08,Morgan10,Mosquera13}.

\item An excess lag in the $U$ band, which samples the Balmer continuum,
suggests that the long-lag problem may be an artifact of
mixing short lags from the disk with longer lags from bound-free 
continuum emission reverberating in the larger BLR
\citep{Lawther18,Korista19,Chelouche19}.
More detailed modeling is needed to see if this
hypothesis can resolve the long-lag problem
and rescue the standard disk theory.
A more radical proposal 
invokes subluminal Alfven-speed signals that trigger
local viscosity enhancements at larger radii
\citep{Sun20}.

\item The time-delay spectrum is
flatter than expected,
$\tau\propto\lambda^{1}$ rather than 
$\tau\propto\lambda^{4/3}$.
This implies a steeper temperature profile
for the accretion disk,
$T\propto r^{-1}$ rather than $r^{-3/4}$.
The best-fit power-law slope is $-0.99\pm0.03$, 
some 7$\sigma$ away from $-3/4$ (Paper~VI).
This might be evidence of nonzero stress at the innermost
stable circular orbit, which can steepen the temperature
profile to a slope of $-7/8$ \citep{Mummery20}.

\item The accretion-disk spectrum, inferred from the
spectrum of the variable component of the light,
is much fainter than predicted using the $T(r)$ profile inferred from 
$\tau(\lambda)$ (Paper~VI). The disk surface seems to have
a higher color temperature, $T(r)$ from the time-delay
spectrum $\tau(\lambda)$, than its brightness temperature,
$T(r)$ from the flux spectrum $F_\nu(\lambda)$.
This low surface brightness and/or high color temperature
is a further challenge to accretion-disk theory.
One possibility is large-grained gray dust obscuring
the AGN, but that would produce a large
mid-infrared excess that is not observed.
Other possibilities are strong local temperature
structures, or azimuthal structures in the disk thickness
casting shadows on the irradiated disk surface.

\item 
The light curve needed to drive continuum reverberations
in the UV and optical differs in detail from the X-ray
light curve (Paper~VI), being smoother and lacking the
rapid variations seen in the X-rays.
\cite{Gardner17} have suggested that the observations
imply that  the
standard inner disk is largely replaced by a geometrically
thick Comptonized region.
Another related possibility is tilting
the inner disk to align with the black hole spin (Paper~VI).

\end{itemize}

These continuum reverberation results pose
serious challenges to the \citet{SS73} accretion-disk theory, 
sparking new thinking on the nature of black hole accretion disks.
The emission-line variations also revealed some unexpected
new phenomena, as follows.

\begin{itemize}

\item There was a significant anomaly in the broad emission line
behavior, the ``BLR Holiday'' (Paper~IV).
The emission lines track the continuum variations
as expected in the first 1/3 of the \storm\ campaign,
but then become fainter than expected
in the latter 2/3, recovering just before the end.
This anomalous period violates the expected 
behavior of emission lines reverberating
with time delays relative to the continuum.
There are also significant changes in line intensity ratios,
suggesting partial covering of a structured BLR,
and/or changes in the shape of the ionizing spectrum.
A plausible interpretation of this BLR Holiday is
that part of the BLR is temporarily obscured to our
line of sight and/or shielded from the ionizing
radiation by a wind outflow, launched from the inner disk,
that can transition between transparent and translucent states
 \citep{Dehghanian19b}.

\item Significant broad and narrow absorption lines
are seen in the UV spectra (Paper~VIII).
The narrow absorption lines exhibit equivalent-width
variations that correlate with the
continuum variations. Here the time delays reflect
recombination times, there being no light travel-time delays
since absorption occurs only along the line of sight.
The inferred density of $\sim10^5$~cm$^{-3}$
and location at $\sim 3$~pc is compatible with clouds
in the narrow-line region (NLR) \citep{Peterson13}.
\end{itemize}

The focus of this paper is an echo-mapping analysis
of the emission-line variations recorded
in the \storm\ data.
Section~\ref{sec:ps} briefly describes the \hst\ and \mdm\ Observatory
spectra and the \ps\
analysis used to improve calibrations, 
extract mean and root-mean-square (rms) spectra,
and the continuum and emission-line light curves.
Section~\ref{sec:barberpole} presents residuals
to the \ps\ fit, including a ``Barber-Pole'' pattern
suggestive of a rotating structure.
In Section~\ref{sec:1d}, we discuss
the linearized echo model and \memecho\ fit to the
emission-line light curves as time-delayed echos of the 
$1150$\,\AA\ continuum light curve, recovering
the one-dimensional delay maps $\Psi(\tau)$
for each line.
To model the anomalous BLR Holiday, 
we extended the \memecho\  model to include slowly-varying line fluxes
in addition to the reverberations modeled as echoes of the driving light curve.
Section~\ref{sec:2d} presents our velocity--delay maps
from \memecho\ analysis of the reverberating emission-line profiles,
exhibiting the clear signature of an inclined Keplerian
disk with a defined outer rim and front/back asymmetry.
Comparisons with previous results are discussed in
Section~\ref{sec:previous}, and Section~\ref{sec:fini}
closes with a summary of the main conclusions.

\section{\ps\ Spectral Decomposition and Calibration Adjustments}
\label{sec:ps}
\typeout{PrepSpec}

Subtle features in the reverberating spectrum
carry the information of interest; thus, echo-mapping analyzes
are sensitive to small calibration errors and
inaccuracies in error-bar estimates.
The first stage of our analysis is therefore to
fit a simple model decomposing the time-resolved spectra
into a mean spectrum plus variable components
each with their own root-mean-square (rms) spectrum and light curve.
\new{For the optical spectra, the} narrow emission-line components are then used
to adjust the photometric calibration and wavelength scale,
and to equalize time-dependent spectral resolution.
The \ps\ code developed and used for this purpose
has been helpful in several previous
studies \citep[e.g.,][]{Grier13}
and is available online\footnote{
http://star-www.st-and.ac.uk/$\sim$kdh1/pub/ps/prepspec.tar.gz}."

\subsection{ \ps\ Spectral Decomposition  }

\begin{figure*}
\begin{center}
\begin{tabular}{cc}
{\bf (a)  \hfill ~ } & {\bf (b) \hfill ~ }
\\
  \includegraphics[angle=270,width=85mm]{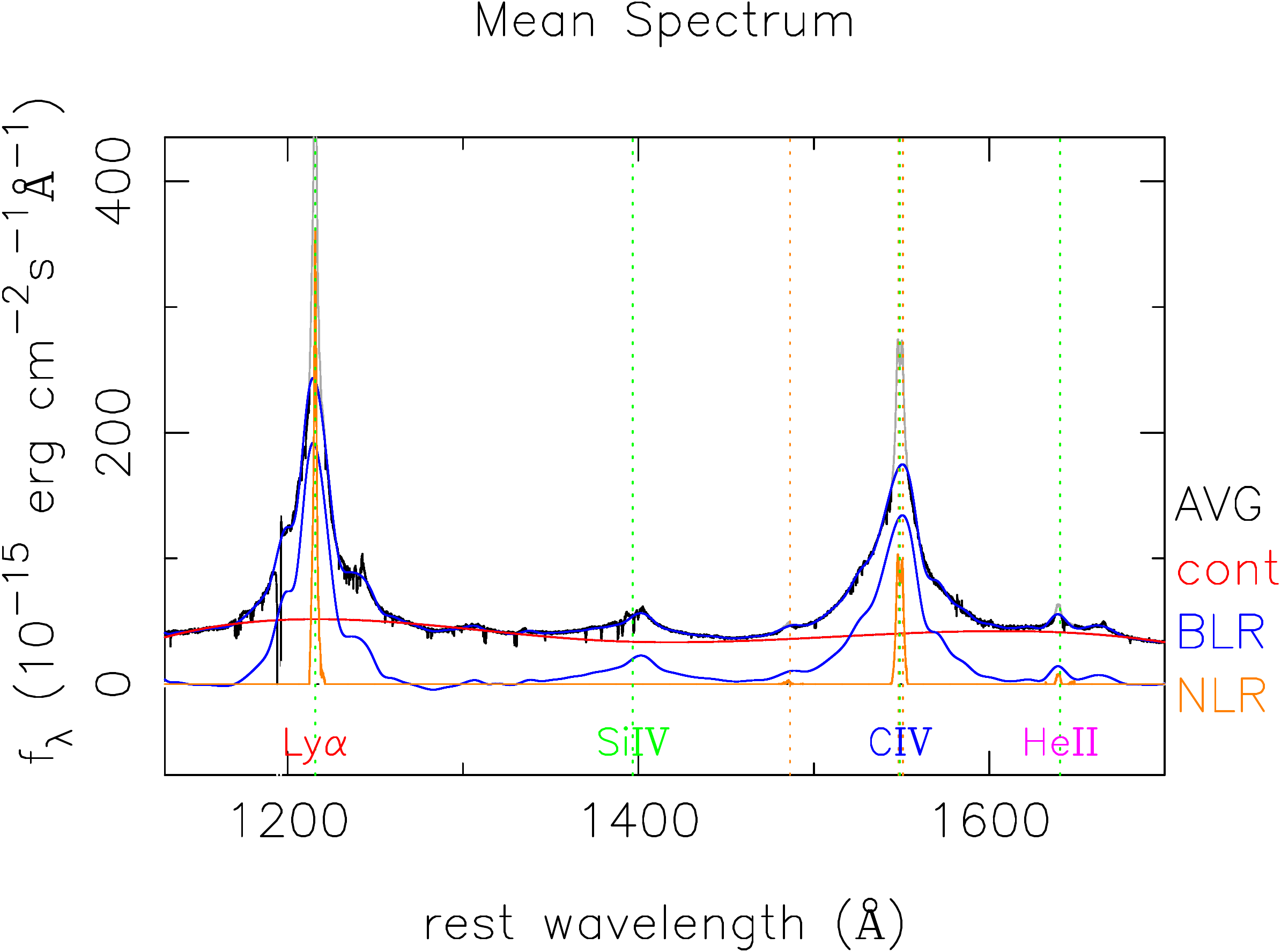}
&
  \includegraphics[angle=270,width=75mm]{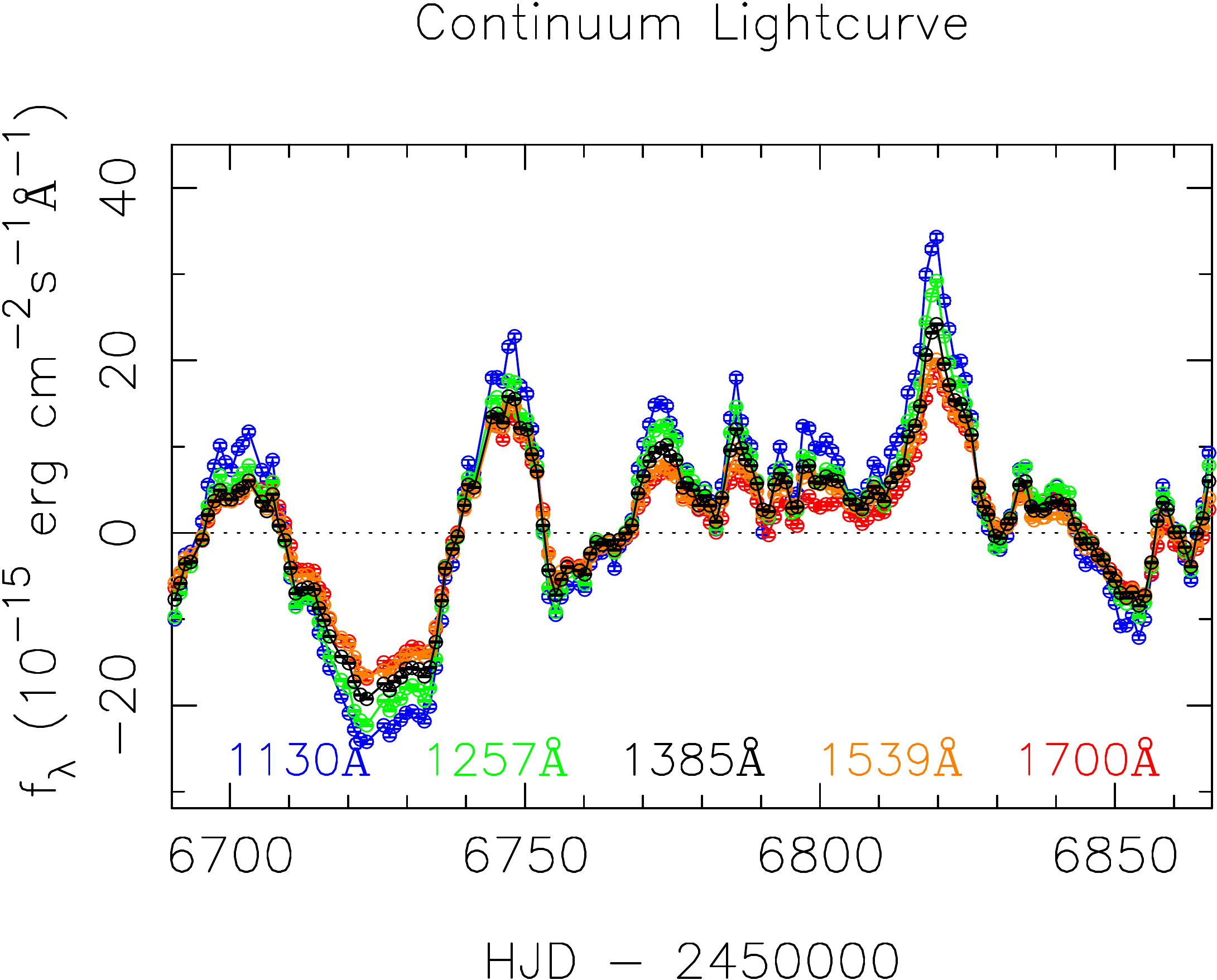}
\\ {\bf (c) \hfill ~ } & {\bf (d) \hfill ~ }
\\
  \includegraphics[angle=270,width=90mm]{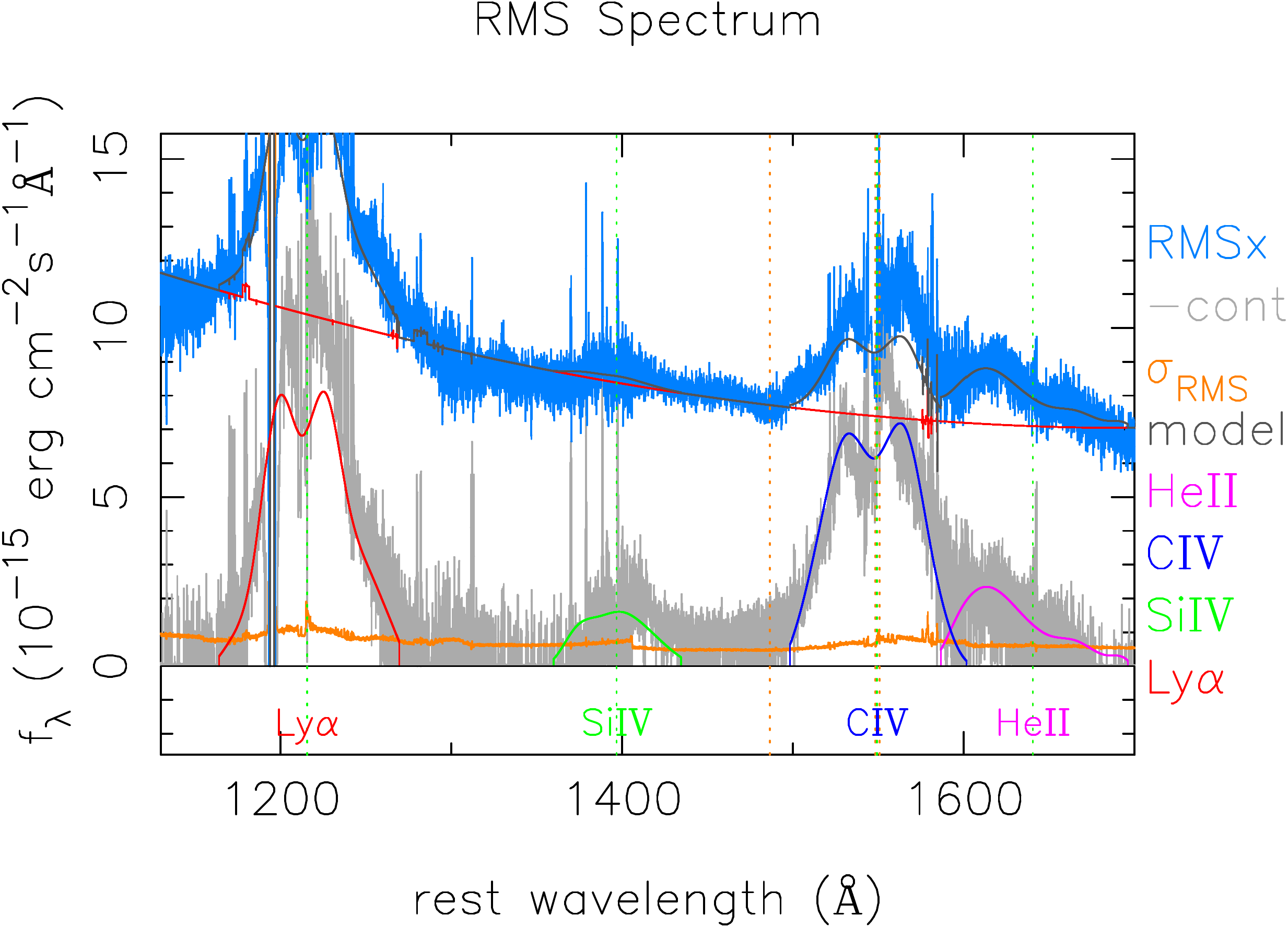}
&
  \includegraphics[angle=270,width=85mm]{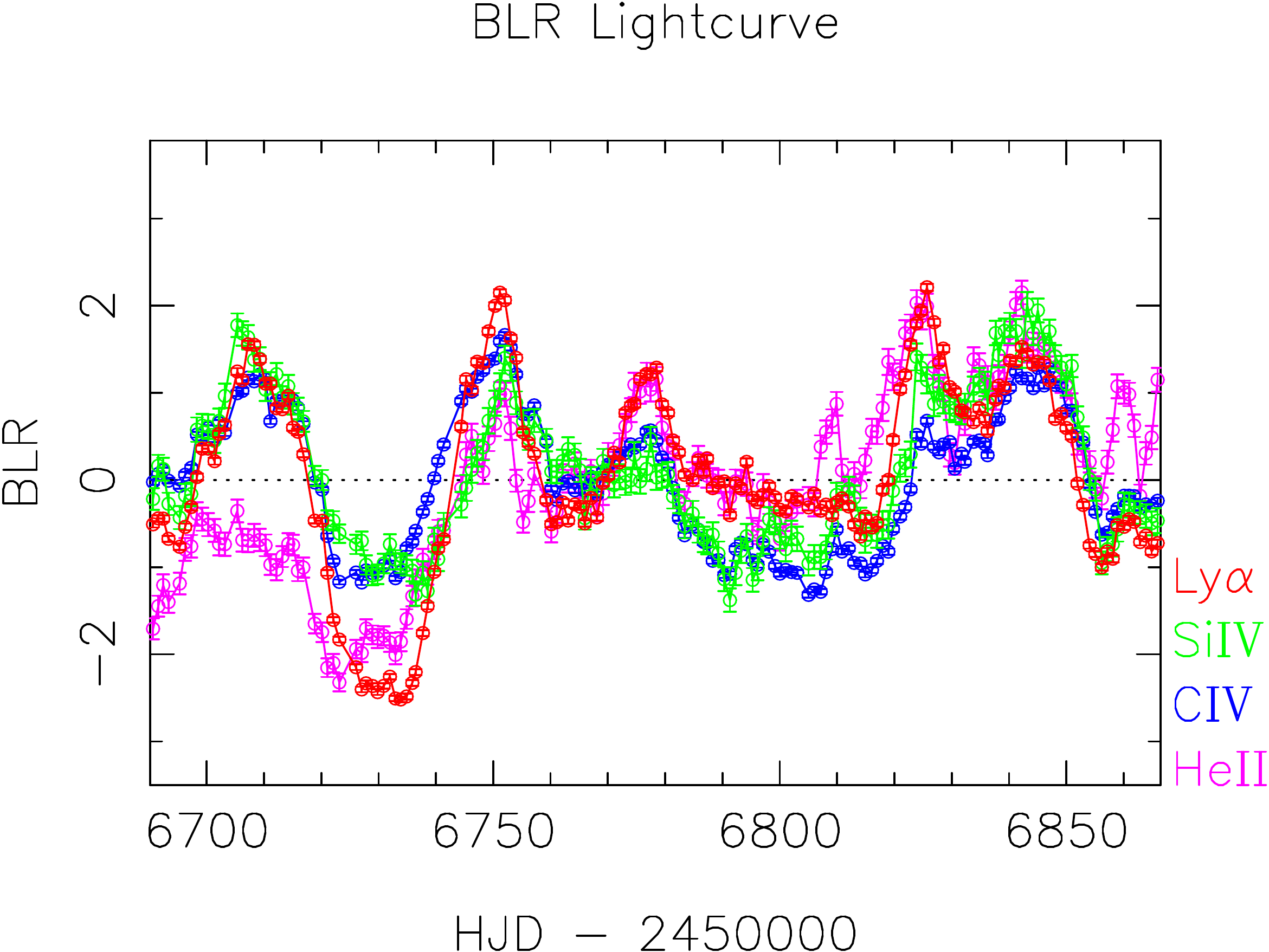}
\end{tabular}
\end{center}

\caption{
Results of the \ps\ fit to the \hst\ data.
 {\bf (a)} The mean spectrum $A(\lambda)$ (gray) is
decomposed into the continuum $\bar{C}(\lambda)$ (red),
the BLR spectrum $\bar{B}(\lambda)$ (blue),
and the NLR spectrum $N(\lambda)$ (orange).
~
{\bf (b)} The continuum light curves, $C(\lambda,t)$,
evaluated at five wavelengths across the spectrum.
The amplitude is larger on the blue end than 
on the red end of the spectrum.
~
{\bf (c)} RMS spectra before and after subtracting the continuum variations (blue
and gray, respectively), and the corresponding uncertainties (yellow).
Also shown are RMS spectra for the fitted model (black),
for the continuum variations $C(\lambda,t)$ (red) and for individual
broad emission lines $B_\ell(\lambda)$ (color coded as indicated).
The blue slope of the continuum variations is evident.
The strong \lya\ and \civ\ lines have double-peaked profiles
in their RMS spectra.
~
{\bf (d)} The BLR light curves $L_\ell(t)$, normalized to a 
median of 0 and a mean absolute deviation of 0.6745
\new{(to match the MAD of an RMS=1 Gaussian).}
\label{fig:pshst}
}
\end{figure*}

The main results of our \ps\ analysis are given in
Fig.~\ref{fig:pshst} for the ultraviolet \hst\ spectra
and in Fig.~\ref{fig:psmdm} for the optical \mdm\ spectra,
where the left column gives the mean and rms spectra 
and the right column the continuum and emission-line light curves.
\ps's  model for spectral variations is
\begin{equation} \label{eqn:abc}
	F(\lambda,t) = A(\lambda)
	+ B(\lambda,t)
	+ C(\lambda,t)
\ ,
\end{equation}
where $A(\lambda)$ is the mean spectrum,
$B(\lambda,t)$ models the broad emission-line variations,
and $C(\lambda,t)$ models continuum variations.
We detail these components below.

\ps\ decomposes the {\it mean spectrum} as
\begin{equation}
	A(\lambda) = N(\lambda) + \bar{B}(\lambda) + \bar{C}(\lambda)
\ ,
\end{equation}
where
$N(\lambda)$ is the NLR spectrum,
$\bar{B}(\lambda)$ is the BLR spectrum,
and $\bar{C}(\lambda)$ is the continuum.
These components are modeled as
 piecewise-cubic spline functions
with different degrees of flexibility: stiff for the continuum, 
more flexible for the BLR, and very loose for the NLR. 
The emission-line components are forced to vanish
outside a range of velocities around the rest wavelength of each line. 
After some experimentation, we set the emission-line windows to
$\pm1500$~\kms for the NLR lines,
$\pm10,000$~\kms for most of the BLR lines,
and  $\pm6000$~\kms\ for \hb.
This decomposition can be used to measure emission-line
strengths, widths, and velocity profiles in the mean spectrum.
However, here we use it mainly to isolate the NLR 
component $N(\lambda)$, which \ps\ uses to
improve the flux and wavelength calibrations.

\ps\ models the {\it continuum variations} 
as low-order polynomials in $\log(\lambda)$,
\begin{equation}
	C(\lambda,t) = \sum_{k=1}^{N_c}
	C_k(t) \, X(\lambda)^k
\ ,
\end{equation}
with $N_c$ coefficients $C_k(t)$ that depend on time.
Here
\begin{equation}
	X(\lambda) = \fracd{ 
	\log{ ( \lambda^2 / \lambda_1\, \lambda_2 ) }
	}{
	\log{ ( \lambda_2 / \lambda_1 ) }
	}
\end{equation}
interpolates linearly in $\log{\lambda}$ from $-1$ to $+1$
over the spectral range $\lambda_1$ to $\lambda_2$.
We adopt cubic polynomials, $N_c=4$, 
to represent the continuum variations in the
\hst\ spectra over the rest-frame wavelengths 1130--1720\,\AA\,
and linear polynomials, $N_c=2$,
for  the optical \mdm\ spectral range 4500--5400\,\AA.
Lower values of $N_c$ leave evident fit residuals
and higher values do not significantly improve the fit.
\ps\ uses the full spectral range to define continuum variations
relative to the mean spectrum, rather than fitting continua to individual
spectra using defined relatively line-free continuum windows.

\ps\ models the {\it broad emission-line variations} as
\begin{equation}
	B(\lambda,t) = \sum_{\ell=1}^{N_\ell}
	\, B_\ell(\lambda) \, L_\ell(t)
\ ,
\end{equation}
thus representing the variable component of each line $\ell$ 
as a fixed line profile $B_\ell(\lambda)$
scaled by its light curve $L_\ell(t)$.
The light curves are normalized to $\left<L_\ell\right>=0$
and $\left<L_\ell^2\right>=1$.
This constraint eliminates degeneracies between the model
parameters, and lets us interpret $B_\ell(\lambda)$
as the rms spectrum of the
variations in line $\ell$.
In the same way as for the mean spectrum,
the rms line profiles $B_\ell(\lambda)$ are  
modeled as piecewise-cubic spline functions, 
and set to 0 outside the BLR window for that line.
~
This separable model for the line variations
assumes that each line has a fixed line profile
that varies in strength with time. Residuals to the \ps\
fit then reveal the evidence for any changes in the line profile.
Such changes contain the information we seek
on the velocity--delay structure
of the reverberating emission-line region, 
and can reveal other interesting phenomena
such as the rotating pattern that we discuss 
in Sec.~\ref{sec:barberpole} below.

\subsection{ UV Spectra from \hst\ }

The UV spectra are the same \hst\ spectra
discussed and analyzed with cross-correlation
methods in Paper~I.
These spectra exhibit several narrow 
absorption systems that interfere with our analysis.
We used the spectral modeling analysis in
Paper~VIII to identify wavelength regions
affected by narrow absorption features and
remove the narrow absorption effects.
The fluxes and uncertainties in these regions are divided by
the model transmission function, restoring to a good approximation the
flux that would have been observed in the absence of
the absorption while also expanding the
error bars to appropriately reflect the lower number of
detected photons.

Similarly, a Lorentzian optical depth profile
provided an approximate fit to
the broad wings of the geocoronal \lya\ absorption.
We divided the observed fluxes and their error bars
by the model transmission, approximately compensating
for the geocoronal \lya\ absorption at moderate optical depths.
The opaque core of the geocoronal line was beyond repair, and we
omit those wavelengths (1214.3--1216.8\,\AA) from our analysis.

The main results of our \ps\ fit to the \hst\ spectra
are shown in Fig.~\ref{fig:pshst}, where the left column
shows the spectral and the right column the temporal
components of the model.
~
In Fig.~\ref{fig:pshst}a,
the mean spectrum $A(\lambda)$ is
decomposed into the NLR spectrum $N(\lambda)$ (orange),
the BLR spectrum $\bar{B}(\lambda)$ (blue),
and the continuum $\bar{C}(\lambda)$ (red).
The BLR spectrum has very strong, broad \lya\ and \civ\ emission
extending to $\pm10,000$~\kms,
with weaker counterparts in \siiv\ and \heii.
As \ps\ failed to robustly separate \lya\ and \nv,
we opted to model the \lya$+$\nv\ blend as a single line.
The NLR spectrum is dominated by \lya\ and \civ\ 
with narrow emission peaks also at \nv\ and \heii.
A few narrow absorption features remain uncorrected
that will not adversely affect our analysis.
~
Fig.~\ref{fig:pshst}b
shows the continuum light curves, $C(\lambda,t)$,
evaluated at five wavelengths across the spectrum.
Continuum variations
with a median absolute deviation (MAD) of 16\%
relative to the continuum in the mean spectrum
are detected with S/N $\approx 500$.
The amplitude is larger at 1130\,\AA\ on the
blue end than at 1700\,\AA\
on the red end of the spectrum.
~
Fig.~\ref{fig:pshst}c shows the rms spectra and 
Fig.~\ref{fig:pshst}d the corresponding BLR light curves.
The blue slope of the continuum variations 
is again evident in the rms spectrum.
The BLR variations are detected with high S/N,
$\sim400$ for \lya\, $\sim300$ for \civ\, $\sim120$ for \heii,
and $\sim80$ for \siiv.
The BLR light curves generally resemble those
of the continuum, but with time delays and other
systematic differences that are distinct
for each line. 
The strong \lya\ and \civ\ lines are single-peaked in the
mean spectrum, but double-peaked 
in the rms spectrum, suggesting that the variations
arise from a disk-like BLR.
Variations are detected in \nv\,$\lambda1240$
on the red wing of \lya, in \siiv\,$\lambda1393$,
and in \heii\,$\lambda1640$.

\subsection{ Optical Spectra from \mdm\ }

\begin{figure*}
\begin{center}
\begin{tabular}{cc}
{\bf (a)  \hfill ~ } & {\bf (c) \hfill ~ }
\\
 \includegraphics[angle=270,width=85mm]{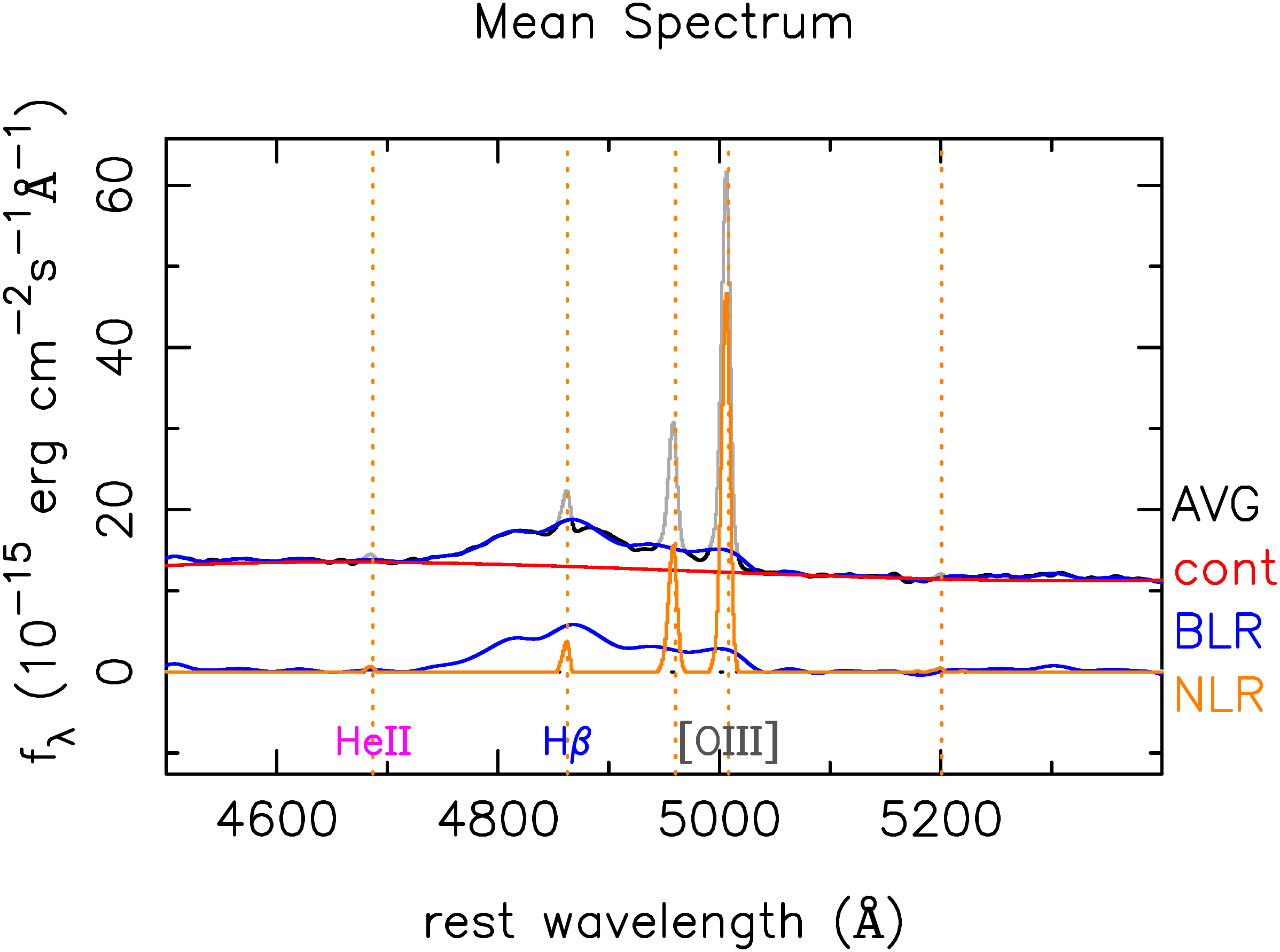}
&
 \includegraphics[angle=270,width=75mm]{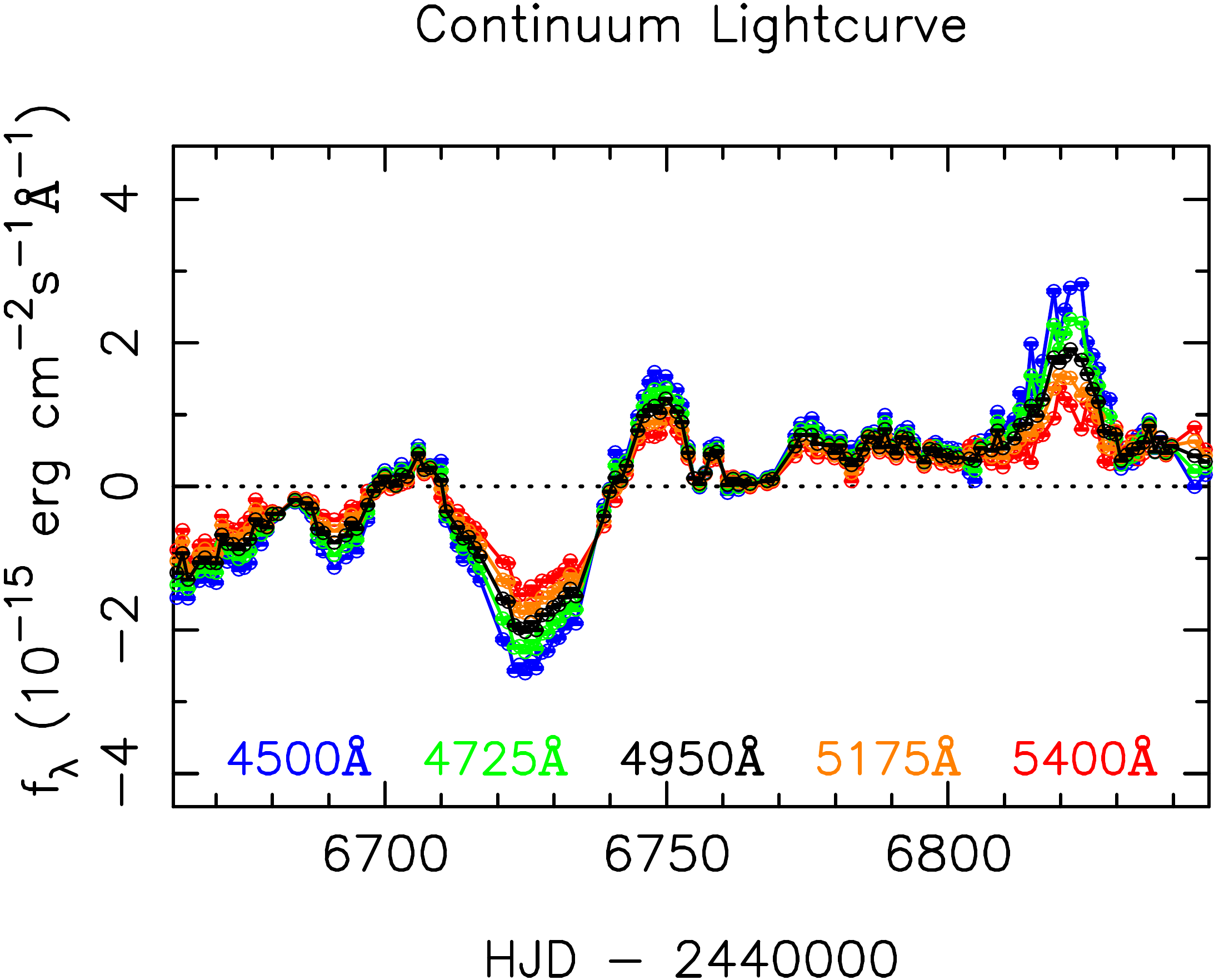}
 \hspace{5mm}
\\
{\bf (b)  \hfill ~ } & {\bf (d) \hfill ~ }
\\
 \includegraphics[angle=270,width=90mm]{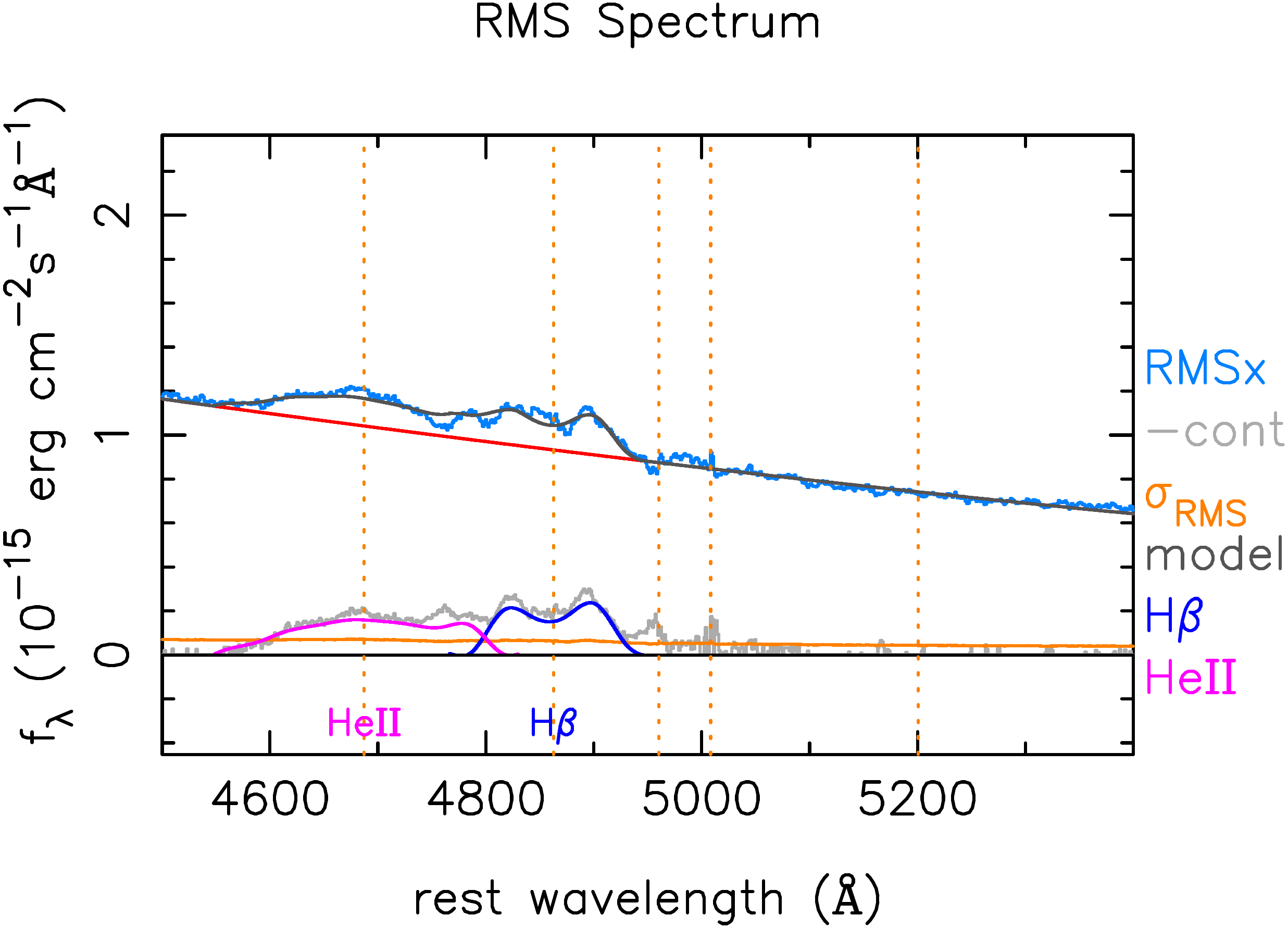}
&
  \includegraphics[angle=270,width=85mm]{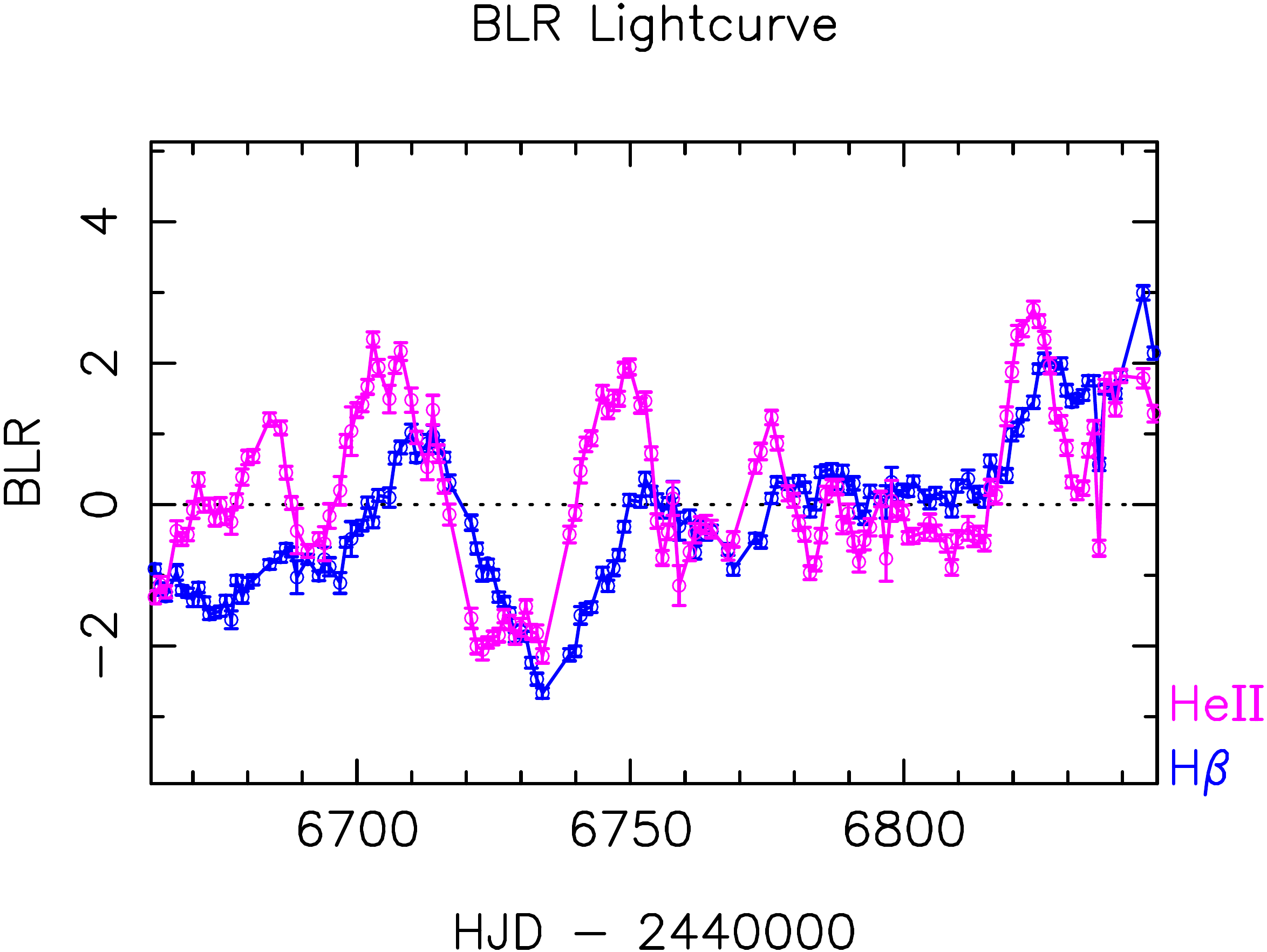}
\end{tabular}
\end{center}
\caption{
As in Fig.~\ref{fig:pshst} but here showing
results of the \ps\ fit to the \mdm\ data covering the optical spectral region 
including the broad \hb\ and  \heii\,$\lambda4686$ and narrow \oiii\ emission lines.
\label{fig:psmdm}
}
\end{figure*}

Optical spectra from the \mdm\ Observatory
were presented and analyzed with a cross-correlation
analysis in Paper~V.
Ground-based
spectra taken at facilities other than \mdm\ were excluded from this
analysis in order to have a consistent and homogeneous dataset taken with the
same instrument, same spectral resolution, and so on.
The ground-based \mdm\ spectra were taken through a $5''$ wide
slit, and extracted with a $15''$ aperture, under
variable observing conditions.
As a result, each spectrum has a slightly
different calibration of flux, wavelength, and spectral resolution.
While these residual calibration errors are most evident
in the regions around narrow emission lines, they
contribute to a smaller extent throughout the spectrum.

To compensate for this, \new{the \ps\ model \new{$M(\lambda,t)$} includes small adjustments to the calibrations:}
\begin{equation}
	M(\lambda,t) = p( t ) \left(
	F
	- \Delta\lambda( t )\,
	\fracd{ \partial F }
	{ \partial\lambda }
	+ \Delta s( t )\,
	\fracd{ \partial^2 F }
	{ \partial\lambda^2 }
	\right)
\ .
\end{equation} 
\new{Here the calibration-adjusted model is $F(\lambda,t)$,
given by Eqn.~(\ref{eqn:abc}),
and the small time-dependent adjustments to the calibration are
parameterized by $p(t)$ to model imperfect photometry, 
$\Delta\lambda(t)$ for small changes to the wavelength scale,
and $\Delta s(t)$ for small changes in the spectral resolution.}
\ps\ models $\ln{p(t)}$ \new{
to ensure } that $p(t)$ remains positive. 
The median of $p(t)$ is set to unity, since typically a minority
of the observed spectra are low owing to slit losses and imperfect
pointing or variable atmospheric transparency.
While \ps\ can model
$\ln{p(t)}$, $\Delta\lambda(t)$, and $\Delta s(t)$
as low-order polynomials of $\log{\lambda}$,
the wavelength dependence of these calibration adjustments was not needed
over the relatively short wavelength span of the \mdm\ data analyzed here.

The main results of our \ps\ analysis of the \mdm\ spectra are shown in
Fig.~\ref{fig:psmdm}. This optical spectral region
includes the broad \hb\ and \heii\,$\lambda4686$
emission lines, and narrow \oiii\ emission lines.
In the mean spectrum, Fig.~\ref{fig:psmdm}a, \hb\ and \oiii\ are strong
but \heii\ is very weak. \hb\ also has a narrow component.
In the rms spectrum, Fig.~\ref{fig:psmdm}b, 
the continuum is bluer than in the mean spectrum,
the \heii\ line is much stronger, and the \oiii\
line is very weak (\new{If well calibrated, this emission line should not vary at all
and thus should not show a signal in the rms spectrum}).
Note that the profile of broad \hb\ is single-peaked in the mean but double-peaked in the rms spectrum.
The continuum and emission-line light curves are shown in Figs.~\ref{fig:psmdm}c
and d, respectively.
Maxima and minima in the \heii\ light curve occur a few days earlier than their
counterparts in the \hb\ light curve.

\subsection{Patterns in Residuals to the \ps\ Fit}
\label{sec:barberpole}
\typeout{barberpole}

\begin{figure*}
\begin{center}
\begin{tabular}{cc}
{\bf (a)  \hfill ~ } & {\bf (b) \hfill ~ }
\\
 \includegraphics[angle=270,width=80mm]{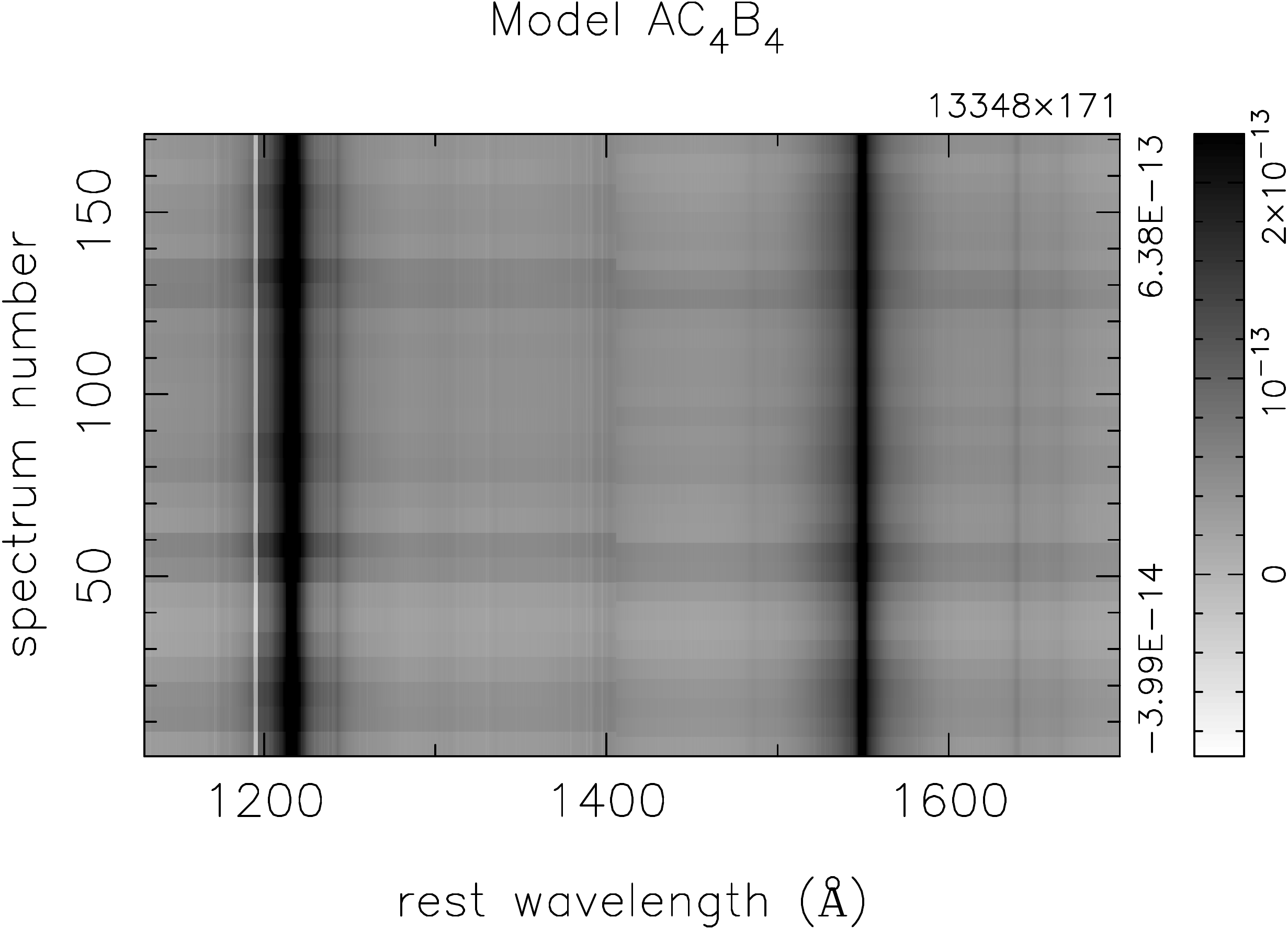}
&
 \includegraphics[angle=270,width=80mm]{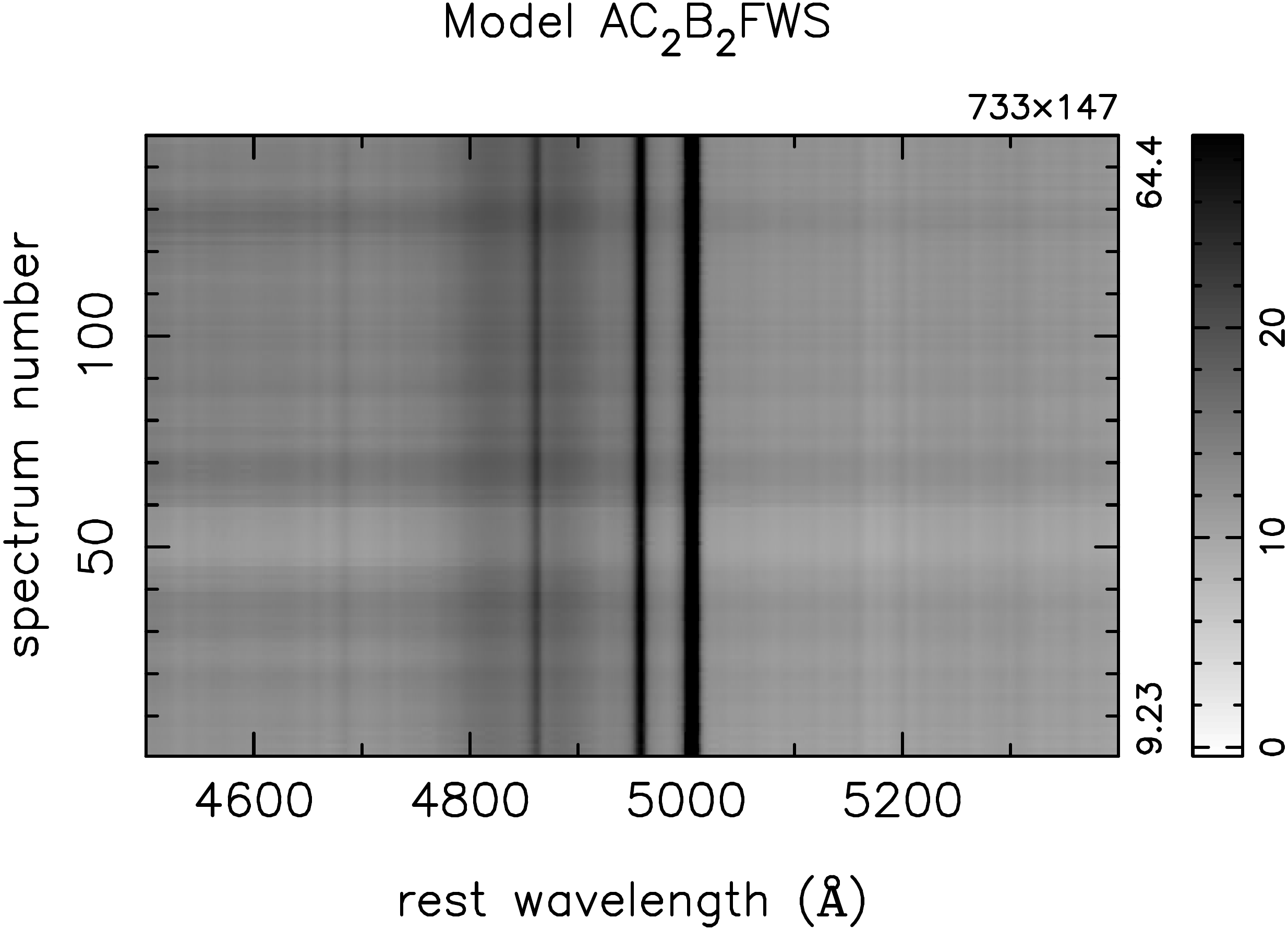}
\\
{\bf (c)  \hfill ~ } & {\bf (d) \hfill ~ }
\\
 \includegraphics[angle=270,width=80mm]{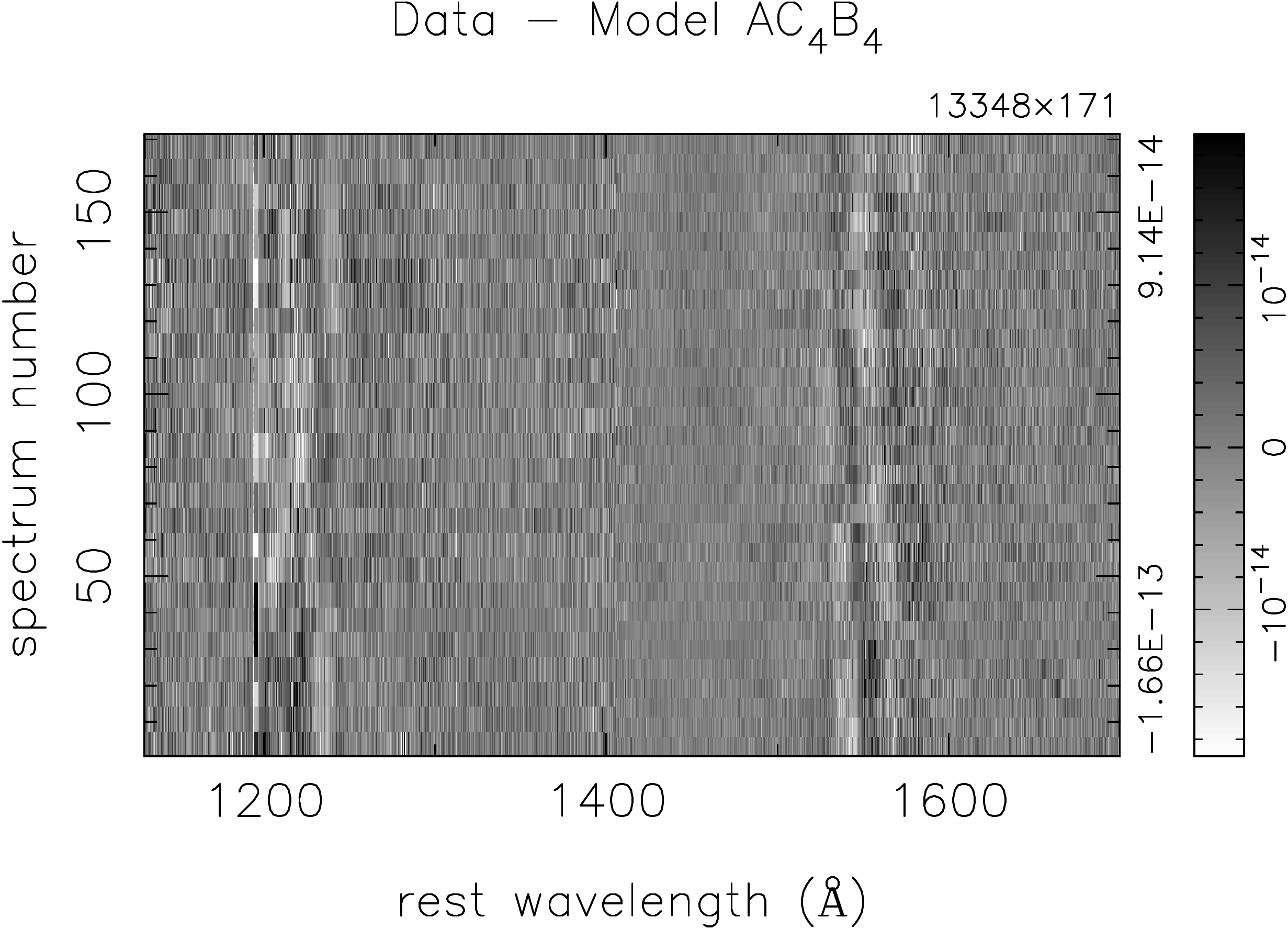}
&
 \includegraphics[angle=270,width=80mm]{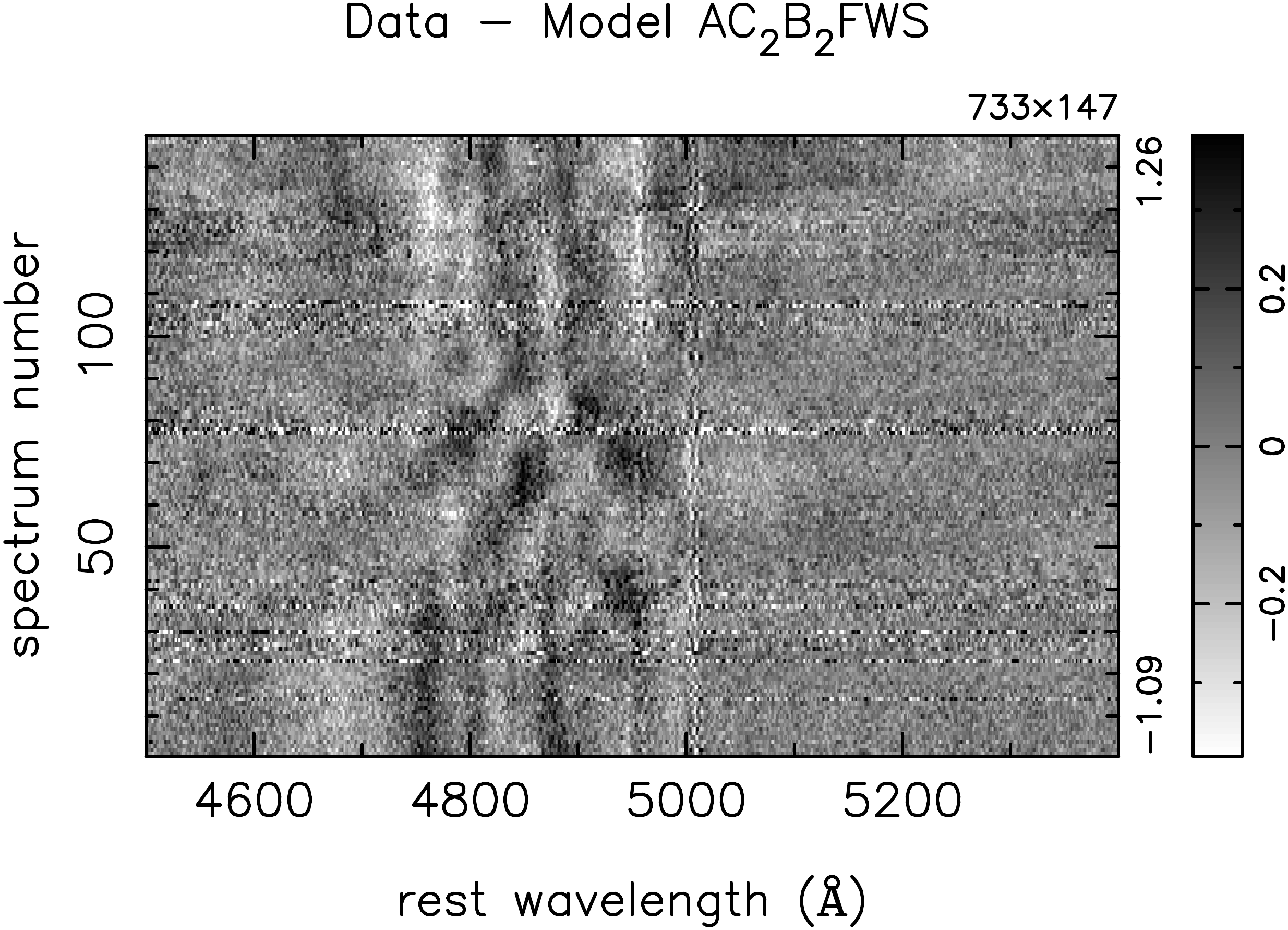}
\\
{\bf (e)  \hfill ~ } & {\bf (f) \hfill ~ }
\\
 \includegraphics[angle=270,width=80mm]{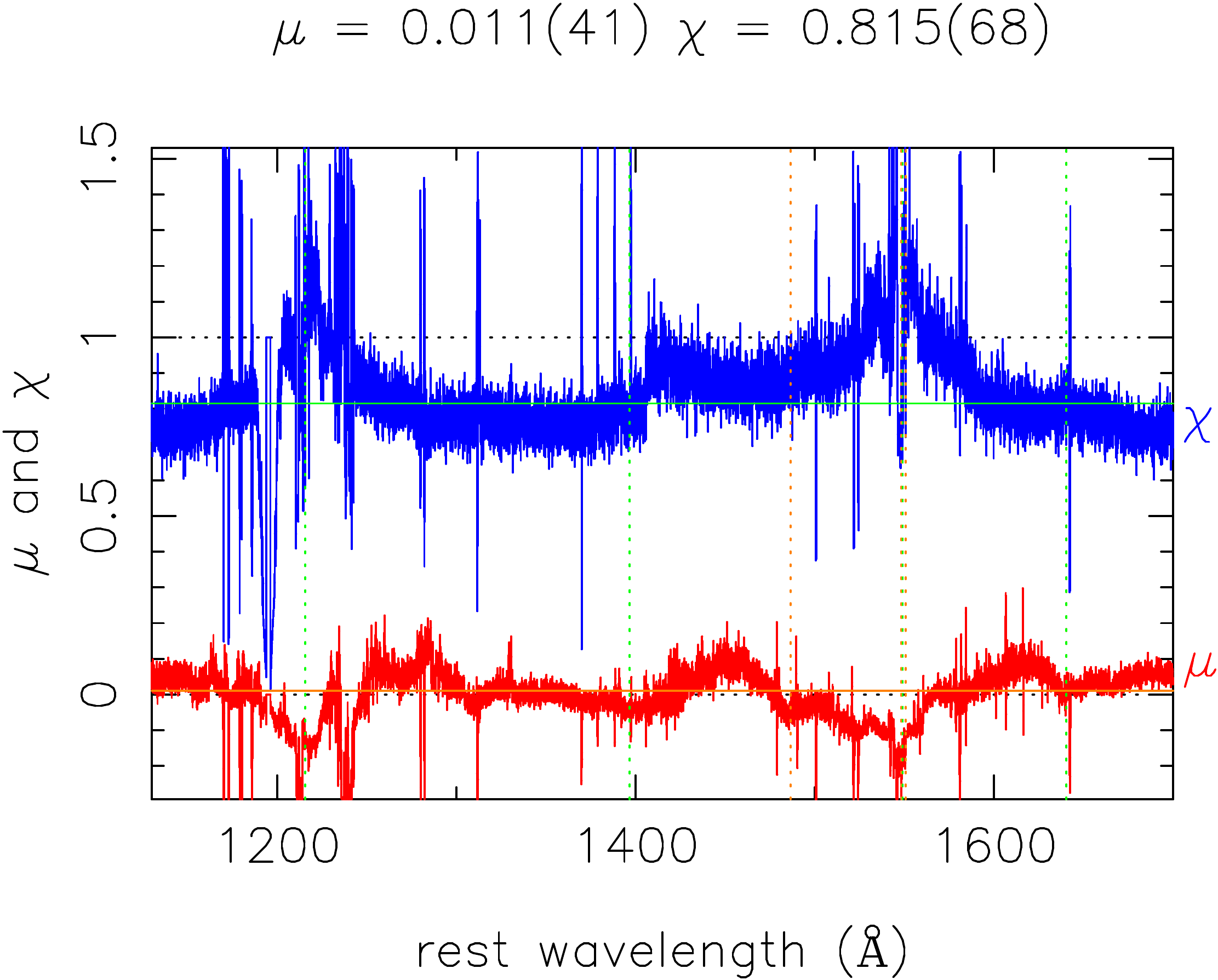}
&
 \includegraphics[angle=270,width=80mm]{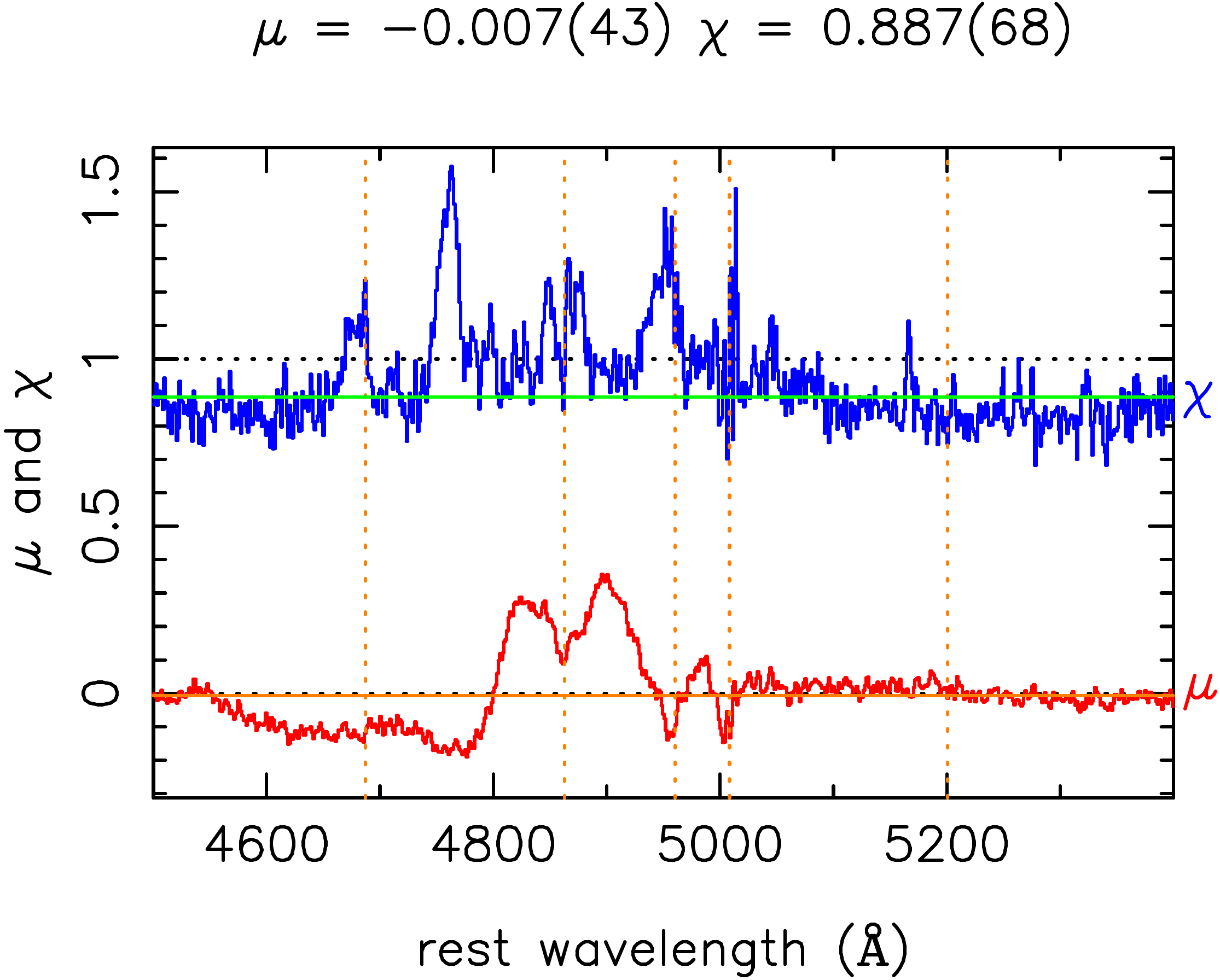}
\\
\end{tabular}
\end{center}
\caption{
Model (a,b) and residuals (c,d) of the \ps\ fit to 
the \hst\ (left) and \mdm\ (right) data.
Lower panel shows the mean (red) and rms (blue) over time
of the normalized residuals for \hst\ (e) and \mdm\ (f).
Note in panel (c) the helical ``Barber-Pole'' pattern of stripes moving from red to blue
across the \civ\ and \lya\ line profiles.
The Model specification key includes components A=average spectrum, C$_n$=continuum polynomial with $n$ parameters,  B$_n$=BLR for $n$ emission lines, and for the \mdm\ data the calibration adjustments F=flux, W=wavelength and S=seeing.
\label{fig:resids}
}
\end{figure*}

Fig.~\ref{fig:resids} presents results of an analysis of
the residuals of the \ps\ fit to the UV \hst\ (left)
and optical \mdm\ (right) spectra.
The \ps\ model assumes for each line
a fixed line profile that changes in normalization only.
The residuals to the \ps\ fit thus present a visualization of
the evidence for variations in the velocity profiles of the emission lines.
They also serve as a check on the success of the absorption-line corrections,
 the calibration adjustments based on the narrow \oiii\ emission lines,
and the accuracy of the error estimates.

The upper panels, Fig.~\ref{fig:resids}a and b, 
present the fitted \ps\ model as a grayscale ``trailed spectrogram,''
with wavelength increasing to the right and time upward.
Here horizontal bands arise from the continuum variations and
vertical bands mark the locations of emission lines.
~
The middle panels, Fig.~\ref{fig:resids}c and d, show
residuals after subtracting the \ps\ model from the observed spectra.
There are acceptably small fine-scale residuals near the strong
narrow \oiii\ lines at 4959\,\AA\ and 5007\,\AA,
indicating the good quality of the calibrations.
In Fig.~\ref{fig:resids}d the evident patterns moving toward the
center of the \hb\ line arise from reverberations affecting
the line wings first and then moving toward the line center.
We find below that these can be interpreted as 
reverberation of \hb-emitting gas with a Keplerian velocity field.
There are also stationary features near 4750\,\AA, 
4880\,\AA, and 4970\,\AA\ that
decrease over the 180~day span of the observations,
indicating a gradual decrease in the emission-line flux with time.

In Fig.~\ref{fig:resids}c the dominant residuals
near the \civ\ line exhibit an intriguing 
helical ``Barber-Pole" pattern 
with stripes moving from red to blue across the line profile. 
This Barber-Pole pattern may be present also
in the \lya\ residuals, but 
less clearly so owing to higher levels of systematic
problems created by the absorption-line corrections
and blending with \nv.
We see no clear evidence of the Barber-Pole pattern in the \hb\ residuals,
where the reverberation signatures are stronger.
The peak-to-trough amplitude of these features in \civ\ is $\pm8$\%
of the continuum flux density --- far too large to be ascribed to 
calibration errors in the \hst\ spectra.

The lower panels, Fig.~\ref{fig:resids}e and f, show the
mean $\mu(\lambda)$ and rms $\chi(\lambda)$ of the {\it normalized}
residuals, scaled by the error bars.
The $\chi(\lambda)$
curves (blue) rise near the emission lines where significant
line-profile variations are being detected, and
level off in the continuum to values below unity,
0.81 for the \hst\ and 0.89 for the \mdm\ spectra.
These low values indicate that rms residuals are smaller 
than expected from the nominal error bars.
For the \memecho\ analysis to follow, we multiply the nominal
error-bar spectra by these factors.

\subsection{ Interpretation of the Barber-Pole pattern}

The Barber-Pole pattern is a new phenomenon in AGN.
\citet{Manser19} report a similar pattern of stripes moving from red to blue
across the infrared \caii\ triplet line profiles arising from
a thin ring or disk of gas orbiting a white dwarf.
The 2~hr period is stable over several years, prompting
its interpretation as due to an orbiting planetesimal
perturbing the debris disk around the white dwarf.

We tentatively interpret the Barber-Pole pattern in \ngc\  as evidence for
azimuthal structure, perhaps caused by the shadows cast by the
vertical structure associated with precessing spiral waves
or orbiting material or streamlines near the base of a disk wind,
that rotate around the black hole with a period of $\sim2$~yr. 
This 2~yr period is estimated based on the impression from Fig.~\ref{fig:resids}c
that the stripes move halfway across the \civ\ profile during the 180~day campaign,
so that 180~days is 1/4 of the period of the rotating pattern.
This is clearly just a rough estimate.
From the velocity--delay maps discussed in Sec.\ref{sec:2d} below,
we infer a black hole mass $\mbh\approx7\times10^7\,\msun$
and a disk-like BLR geometry extending from 2 to 20 light days
with an inclination $i \approx 45^\circ$.
A 2~yr orbital period then occurs at $R/c \approx 4$~days, or $R \approx 1000~G\,\mbh/c^2$,
compatible with the inner region of the BLR.
The corresponding Kepler velocity is $V=\sqrt{G\,\mbh/R} \approx 9000$~\kms,
and this projects to $V\,\sin{i} \approx 7000$~\kms\ for $i\approx45^\circ$.
These rough estimates are compatible with the observed velocity amplitude of the
Barber-Pole stripes in the \civ\ residuals.

The effect must be stronger
on the far side of the disk to produce Barber-Pole features that move from red to blue
across the line profile, and weaker on the near side, where they
would be seen moving from blue to red.
This front-to-back asymmetry might be due to a bowl-shaped BLR geometry, 
so that the near side of the BLR disk is strongly foreshortened. 
However, the velocity--delay maps discussed in Sec.\ref{sec:2d}
indicate that the response is stronger on the near side than on the far side
of the disk.
Alternatively, if the inner disk is tilted toward us, perhaps due to a misaligned black hole spin,
then material orbiting there could rise above the outer-disk plane,
to cast shadows on the far side of the outer disk,
and then dip below the plane to avoid casting shadows on the near side of the outer disk.

Detailed modeling beyond the scope of this paper may test the viability of 
these and other interpretations. Further monitoring of \ngc\ with \hst\ may be 
helpful to determine if the Barber-Pole phenomenon is stable or transient, 
whether its period is stable or changing, and whether the stripes always go
from red to blue or sometimes from blue to red across the \civ\ profile.

\section{ \memecho\ Analysis: Velocity--Delay Mapping }
\label{sec:1d}
\typeout{1d}

\new{Our echo-mapping analysis is performed with the \memecho\ code,
which is described in some detail by \cite{Horne94}. 
Its ability to recover velocity--delay maps from simulated \hst\ data
is demonstrated \citep{Horne04} and it has
recently been subjected to blind tests \citep{Mangham19}.
We outline below the assumptions and methodology,
and then present and discuss the results of our
\memecho\ analysis of the \hst\ and \mdm\ data on \ngc.}

 Echo mapping assumes that a compact source of ionizing radiation
is located at or near the center of the accretion flow. 
Photons emitted here shine out into the surrounding region,
causing local heating and ionization of gas which
then emits a spectrum characterized by emission lines
as it cools and recombines.
Reprocessing times are expected to be short and dynamical times long
compared to light-travel times.
As distant observers, we see the response from each
reprocessing site with a time delay $\tau$ from the light-travel time
and a Doppler shift $v$ from the line-of-sight velocity.
Thus, the reverberating emission-line spectrum encodes information about the
geometry, kinematics, and ionization structure of the accretion flow ---
to be more specific, that part of the flow that emits the reverberating
emission lines.

To decode this information, we interpret the observed spectral variations
as time-delayed responses to a driving light curve.
By fitting a model to the reverberating spectrum $F(\lambda,t)$,
we reconstruct a two-dimensional wavelength-delay map $\Psi(\lambda,\tau)$.
This effectively slices up the accretion flow on isodelay surfaces,
which are paraboloids coaxial with the line of sight with a
focus at the compact source.
Each delay slice gives the spectrum of the response,
revealing the fluxes and Doppler profiles of emission lines
from gas located on the corresponding isodelay paraboloid.
The resulting velocity--delay maps $\Psi(v,\tau)$ provide
two-dimensional images of the accretion flow, one for each emission line,
resolved on isodelay and isovelocity surfaces.

\subsection{Linearized Echo Model }

The full spectrum of ionizing radiation is not observable, and so
an observed continuum light curve, $C(t)$, is adopted as a proxy. 
At each time delay $\tau$, the responding emission-line light curve 
$L(t)$ is
then some nonlinear function of the continuum light curve $C(t-\tau)$
shifted to the earlier time $t-\tau$.
In addition, the observed line and continuum fluxes
include constant or slowly varying background contributions from
other light sources,
such as narrow-line emission and starlight from the host galaxy.
To model these backgrounds and account for the nonlinear BLR responses,
\memecho\ employs a {\it linearized} echo model, 
with reference levels $C_0$ for the continuum and $L_0$ for the line flux,
and a tangent-curve approximation to variations around these reference levels. 
Thus the continuum light curve $C(t)$ is decomposed as
\begin{equation}
	C(t) = C_0 + \Delta C( t )
\ ,
\end{equation}
and the emission-line light curve
\begin{equation}
	L(t) = L_0
	+ \int \, \Psi( \tau ) \, \Delta C( t - \tau) \, \dd\tau
\ ,
\end{equation}
is a convolution of the continuum variations with a delay map $\Psi(\tau)$,
giving the one-dimensional delay distribution of the emission-line response.
We find that this linearized echo model fails to provide a good fit to the \ngc\ data.
We therefore generalize the model to allow a time-dependent echo background level,
$L_0(t)$. {\bf This extension is straightforward.}

\subsection{ Maximum Entropy Regularization}
\new{
Maximum entropy regularization keeps the model light curves
$C(t)$ and $L_0(t)$
and the delay maps $\Psi(\tau)$
positive and ``as smooth as possible'' while fitting the data.
Referring to these functions generically as $p(t)$, the entropy is
\begin{equation}
	S(p) = \sum_t w(t) \, \left[ p(t) - q(t) - p(t) \,\ln{ \left( p(t) / q(t) \right) } \right] \ ,
\end{equation}
measured with weights $w(t)$ and
relative to a default function $q(t)$.
We obtain  $q(t)$ by Gaussian smoothing of $p(t)$,
with a full width at half-maximum (FWHM) of 1, 2, and 4~days
for the driving light curve, the delay map, and the
echo background, respectively.
These choices control the flexibility of the functions.
The weights $w(t)$ provide additional control on
relative flexibility among the 3 functions.

For fits to reverberating spectra, the \memecho\ model simply adds
a wavelength dimension to the echo light curve, $L(t) \rightarrow L(\lambda,t)$,
to the response distribution, $\Psi(\tau) \rightarrow \Psi(\lambda,\tau)$,
and to the background variations, $L_0(t) \rightarrow L_0(\lambda,t)$.
These 2-dimensional functions are then regularized with the entropy
defined relative to default functions that average in both directions.

The \memecho\ fit is accomplished by iteratively adjusting
the  functions $p$ to minimize
\begin{equation}
	 Q \left( p, D \right) = \chi^2 \left( p, D \right) - \alpha \, S \left( p \right) \ .
\end{equation}
Here $\chi^2\left( p, D \right)$ quantifies the ``badness of fit" to the $N$ data $D$, 
assuming Gaussian noise with known error bars.
The Lagrange multiplier $\alpha$ controls
the tradeoff between fitting the data (small $\chi^2$) and being simple (large $S$).
In practice, $\alpha$ is initially large and a series of converged fits is constructed
with decreasing $\chi^2$ and increasing $S$, stopping when the fit 
is judged to be satisfactory ($\chi^2/N\approx1$) and the model not overly complex.
}

\subsection{Delay Maps $\Psi(\tau)$ for \ngc\ }

\begin{figure*}
\begin{center}
 \includegraphics[angle=0,width=130mm]{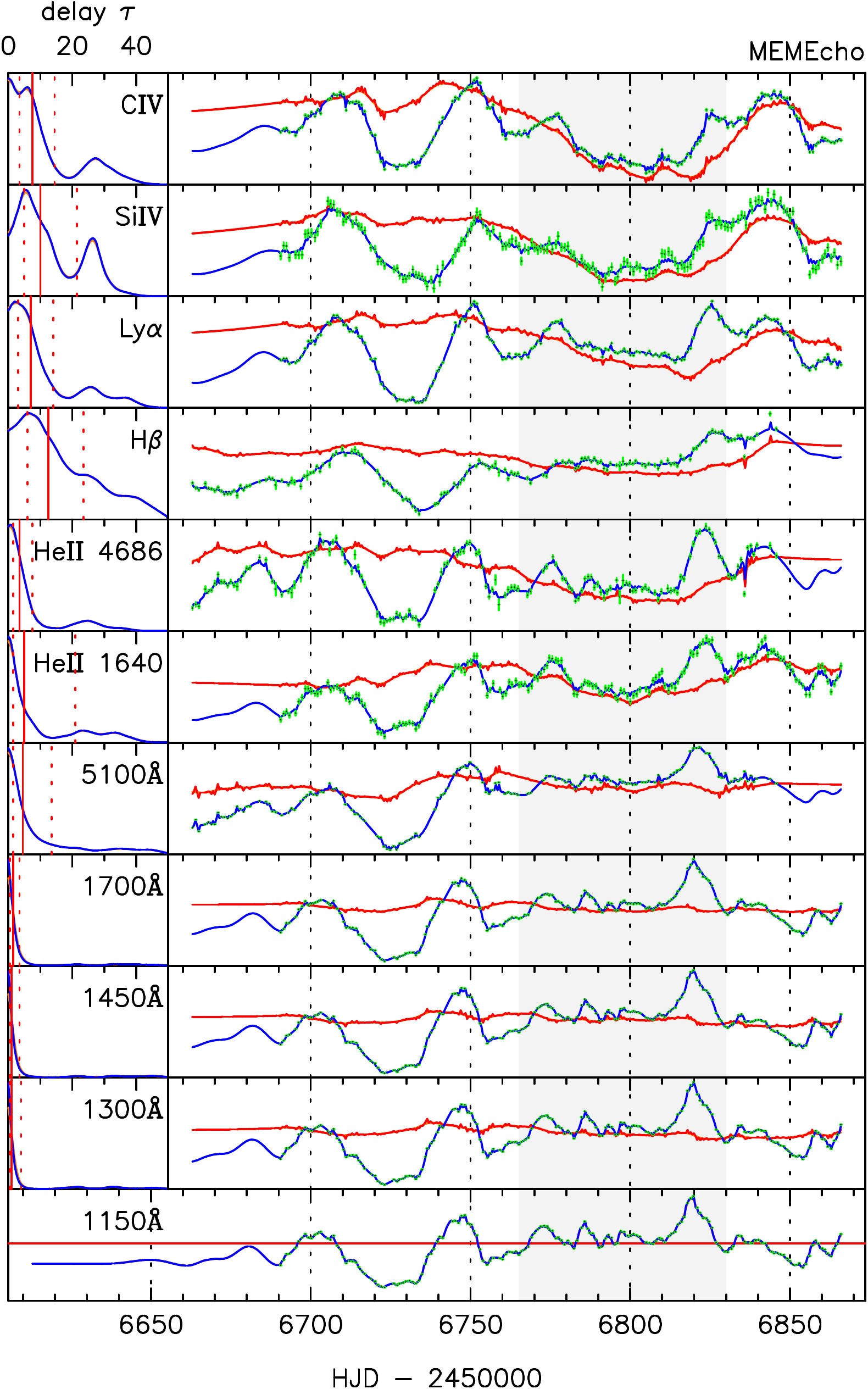}
\caption{
\label{fig:1d}
\memecho\ fits to five continuum and six emission-line light curves of \ngc.
The driving light curve (bottom panel) is the 1150\,\AA\ continuum light curve
with a reference level (red line) at the median of the 1150\,\AA\ continuum data.
Above this are ten echo light curves (right) and corresponding delay maps (left),
comparing the light-curve data (the black points with green error bars) 
and the echo model (blue curves) with slow background variations (red curves).
The echos (bottom up) are four continuum light curves,
at 1300\,\AA, 1450\,\AA, and 1700\,\AA\ from the \hst\ spectra
and at 5100\,\AA\ from the \mdm\ spectra,
then six reverberating emission lines 
(\heii\,$\lambda1640$, $\lambda4686$,
\hb, \lya, \siiv, and \civ).
On each delay map in the left column,
the median delay is marked by a vertical line,
flanked by vertical dashed lines for the quartiles of the delay distribution.
The \memecho\ fit accounts for much of the light-curve structure
as echos of the driving light curve, but requires
significant additional variations  (red curves).
\new{The gray shaded region indicates the timespan of the ``BLR Holiday'' identified in Paper~IV.}
}
\end{center}
\end{figure*}

Figure \ref{fig:1d} shows the results of our \memecho\ fit to
five continuum and six emission-line light curves of \ngc.
The light-curve data are from the \ps\ analysis 
of the \hst\ and \mdm\ spectra, described in Sec.~\ref{sec:ps}.
\memecho\ fits all light curves simultaneously,
recovering a model for the driving light curve $C(t)$,
and for each echo light curve a delay map $\Psi(\tau)$
and a background light curve $L_0(t)$.
The driving light curve $C(t)$ (bottom panel of Fig.~\ref{fig:1d})
is the 1150\,\AA\ continuum light curve, with
the reference level $C_0$ (red line) set at the median of the 1150\,\AA\ continuum data.
Above this are ten echo light curves (right) and corresponding delay maps (left),
where the light-curve data (black points with green error bars) can be directly
compared with the fitted model (blue curves).
We model four continuum light curves,
at 1300\,\AA, 1450\,\AA, and 1700\,\AA\ from the \hst\ spectra
and at 5100\,\AA\ from the \mdm\ spectra,
as echos of the 1150\,\AA\ continuum.
The reverberating emission lines are \heii\,$\lambda1640$ and \heii\,$\lambda4686$,
then \hb\ and \lya, and finally \siiv\ and \civ.
The \memecho\ fit accounts for much of the light-curve structure
as echos of the driving light curve, but requires
significant additional variations $L_0(t)$  (red curves),
\new{particularly during the BLR Holiday indicated by gray shading
in Fig.~\ref{fig:1d}.}

The fit shown in Fig.~\ref{fig:1d} requires $\chi^2/N=1$ separately
for the driving light curve and for each of the echo light curves,
where there  are $N=171$  and 
$N = 147$ data points for the \hst\ and \mdm\ light curves, respectively.
The model light curves (delay maps) are evaluated on 
a uniform grid of times (delays) spaced by $\Delta t=0.5$~days,
linearly interpolated to the times of the observations.
The delay maps span a delay range 0--50~days.

The delay maps $\Psi(\tau)$ are of primary interest because they
indicate the radial distributions from the central black hole
over which the continuum and emission lines are responding
to variations in the driving radiation.
The continuum light curves exhibit highly correlated variations
that are well fit by exponential delay
distributions strongly peaked at $\tau=0$.
The median delay, increasing with wavelength, 
is $\sim1$~day at 1700\,\AA\ and $\sim5$~days at $5100$\,\AA.
The echo background has only small variations,
indicating that the linearized echo model is a very
good approximation for the continuum light curves.

The emission-line light curves require more extended delay
distributions and larger variations in their background levels.
The background variations are similar, but not identical,
for the six emission-line light curves.
The two \heii\ light curves require tight delay distributions
peaking at $\tau=0$, with half the response inside $\sim5$~days
and 3/4 inside 10~days,
and some low-level structure at 20--40~days.
The background light curves have a ``slow wave'' with a 100-day timescale,
somewhat different for the two lines,
and smaller amplitude 10-day structure.
The slow-wave background for \heii\,$\lambda1640$ is rising 
from HJD~6690 to 6750 
(really HJD$-$2,450,000),
while that for \heii\,$\lambda4686$ is more constant.
Both backgrounds then decline to minima around HJD 6800 and then 
rise until 6840.
The constant background for \heii\,$\lambda1640$ prior to HJD 6690
and for \heii\,$\lambda4686$ after HJD 6850 is not
significant since there are no data during these intervals.

The \hb\ response exhibits the most extended delay distribution,
with a peak at 7~days, half the response inside 14~days,
3/4 inside 23~days, 
minor bumps at 25 and 40~days, and falling to 0 at 50~days.
The need for this extended delay map is evident in
the \hb\ light curve, for example to explain
the slow \hb\ decline following peaks at HJD 6705 and 6745.
The \lya\ response is more confined than \hb\, with a peak at 3~days,
half the response inside 7~days, 3/4 inside 15~days,
and bumps at 26 and 35~days.
\siiv\ and \civ\ are similar, with peaks at 5 and 7~days, respectively.

The slow-wave backgrounds \new{$L_0(t)$} for all these lines fall 
slowly from HJD 6740 to 6820 and then rise more rapidly to a peak at HJD 6840.
This \new{corresponds approximately to} the anomalous BLR Holiday period discussed in Paper~IV,
\new{ indicated by gray shading in Fig.~\ref{fig:1d}, }
during which the emission lines became weak relative to the
continuum.
Note also a smaller dip from HJD 6715 to 6740 that serves to deepen
the emission-line decline between the two peaks,
particularly for \civ.
A small peak near HJD 6810 accounts for emission-line peaks
in \civ, \siiv, and \heii\,$\lambda1640$ that
have no clear counterpart in the continuum light curves.

Note that the model and background light curves (blue and red curves in Fig.~\ref{fig:1d})
exhibit numerous small spikes in addition to smoother 100~day and
10~day structure. These spikes correspond to data points
that are too high or too low, relative to their
error bars, to be fit by the smooth
default light curve that maximizes the entropy.
The largest offender is a low point in the \hb\ and
\heii\,$\lambda4686$ light curves near HJD 6837,
which likely represents a calibration error.
These outliers could seriously damage the delay maps.
The spikes are less prominent if we relax the fit
to a higher $\chi^2/N$, but then the fit 
to the relatively low S/N
\siiv\ light curve is less satisfactory.
Fortunately, because our model has time-dependent backgrounds
that can develop sharp spikes where required,
the delay maps remain relatively smooth and insensitive
to these outliers.

\section{ Velocity--Delay Maps }
\label{sec:2d}
\typeout{2d}

\begin{figure*}
\begin{center}
 \includegraphics[angle=270,width=140mm]{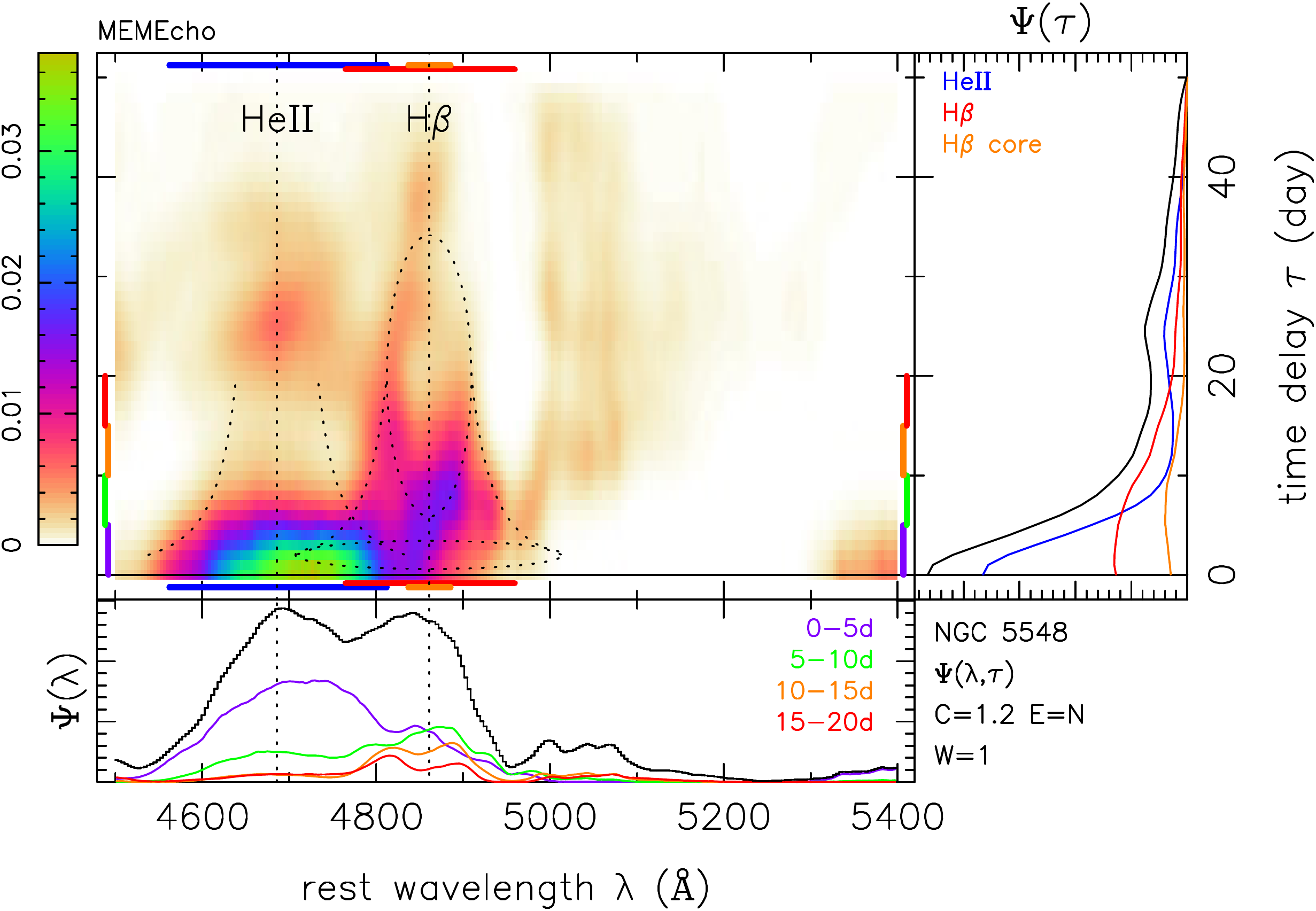}
\caption{ \label{fig:2dmdm}
Two-dimensional wavelength--delay map $\Psi(\lambda,\tau)$
reconstructed from the \memecho\ fit to the optical spectra from \mdm.
Delays are measured relative to the 1150\,\AA\ continuum light curve.
Panels below and to the right of the map
give the projected responses $\Psi(\lambda)$ and $\Psi(\tau)$.
\new{Here the black curve is the full response and the colored curves
are for the wavelength or delay ranges indicated by the correspondingly 
colored bars in the margins of the map.}
\new{Dotted} curves show the envelope around each line
inside which emission can occur from a Keplerian disk
inclined by $i=45^\circ$ orbiting a black hole of mass 
$\mbh=7\times10^7\,\msun$.
\new{The ellipses shown for \hb\ correspond to Keplerian disk orbits at $R=2$ and 20~light days.}
}
\end{center}



\begin{center}
 \includegraphics[angle=270,width=140mm]{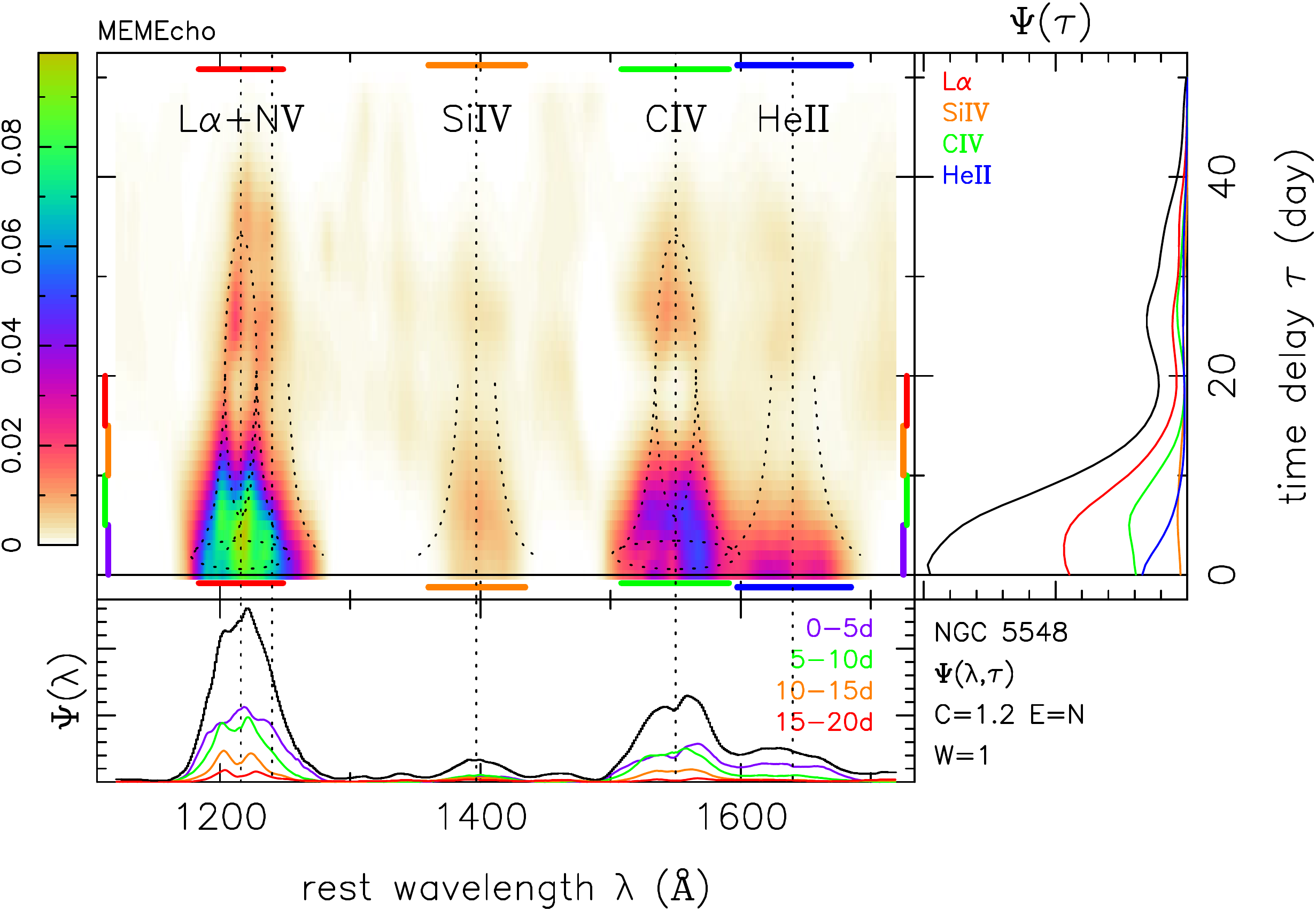}
\caption{ \label{fig:2dhst}
\new{ As in Fig.~\ref{fig:2dmdm}, showing the}
two-dimensional wavelength--delay map $\Psi(\lambda,\tau)$ reconstructed
from the \memecho\ fit to the UV spectra from \hst.
\new{The ellipses shown for \lya\ and \civ\
correspond to Keplerian disk orbits at $R=2$ and 20~light days.}
}
\end{center}
\end{figure*}

\begin{figure*}
\begin{center}
 \includegraphics[angle=0,width=150mm]{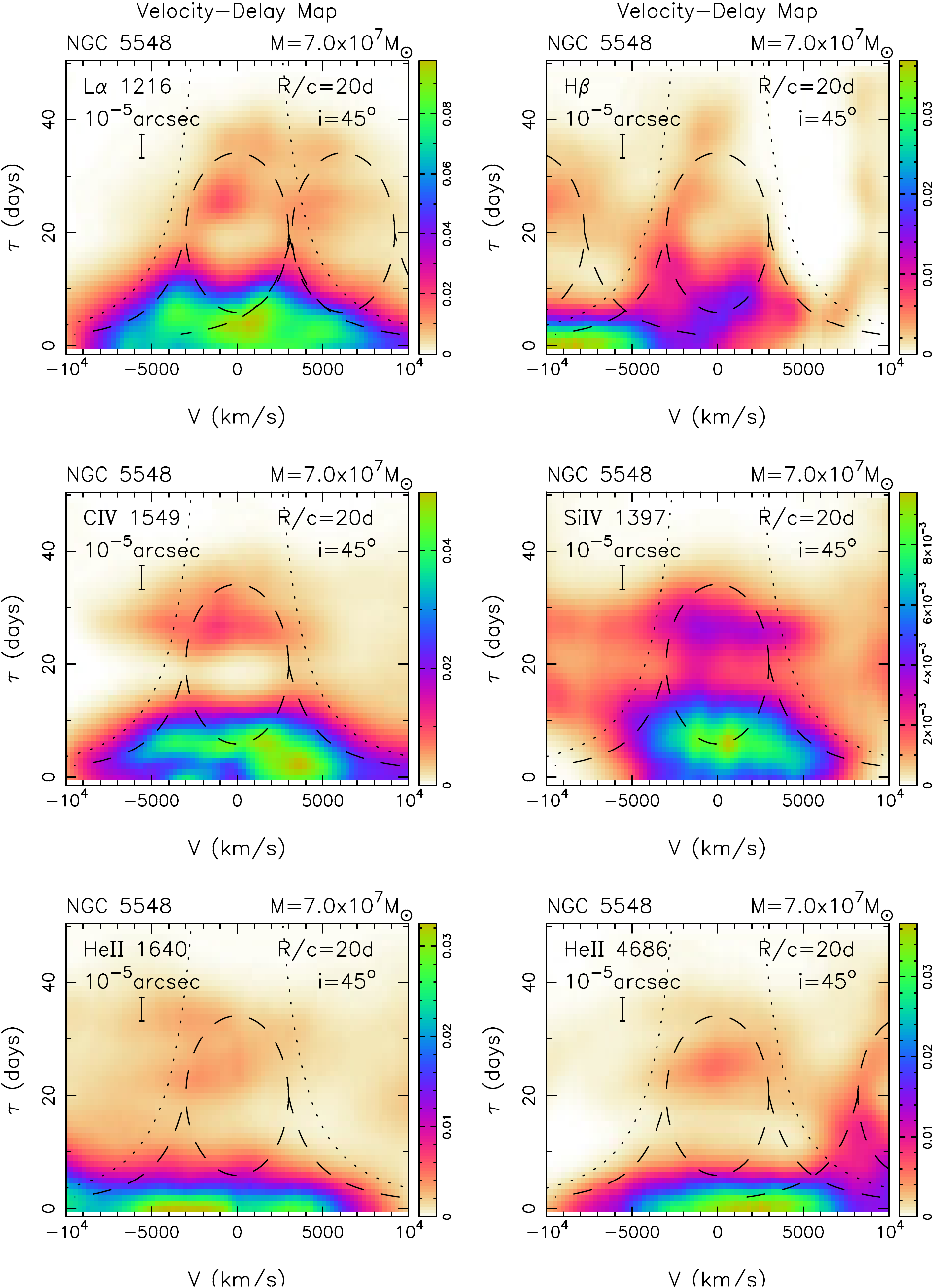}
\caption{ \label{fig:2d}
Velocity--delay maps $\Psi(v,\tau)$ reconstructed
from the \memecho\ fit to the  \hst\
and \mdm\ spectra.
Delays are measured relative to the 1150\,\AA\ continuum light curve.
\new{ Velocities are measured relative to the rest wavelength of the indicated line.}
Dashed \new{black} 
curves show the virial envelope around each line
inside which emission can occur from a Keplerian disk
inclined by $i=45^\circ$ orbiting a black hole of mass 
$\mbh=7\times10^7\,\msun$.
\new{The dashed ellipse corresponds to a circular Keplerian orbit at the outer disk rim radius $R/c=20$ days.}
\new{Note that the red wing of \lya\ is blended with \nv\, and 
the blue wing of \hb\ is blended with \heii\ $\lambda4686$.}
\new{A scale bar corresponding to 10 micro-arcseconds is shown in each panel.}
}
\end{center}
\end{figure*}

Velocity--delay maps project the information that is coded in the reverberating
emission-line profile onto a 2-dimensional map of the 
6-dimensional position-velocity phase space of the BLR gas.
While this is incomplete information, an ordered velocity field
can have an easily recognizable signature in the velocity--delay map;
some examples are shown by \citet{Welsh91}.
A signature of inflowing gas is short delays on the red wing
of the velocity profile, and a wide range of delays on the blue side.
Outflowing gas has a similar but reversed signature.
An orbiting ring of gas at radius $R$ maps into an ellipse on the velocity--delay plane,
centered at $\tau=R/c$ and extending over $(R/c)(1\pm\sin{i})$,
allowing the identification of $R$ and $i$.
A Keplerian disk superimposes these ellipses to form
a ``virial envelope'' that can be used to infer $V\,\sin{i}$ at each $R$.
Assuming $V=\sqrt{G\,\mbh/R}$, this gives $\mbh/\sin^2{i}$.
Thus, a sufficiently crisp velocity--delay map can be
read to infer the general nature of the flow in the BLR, and
several specific parameters of the geometry and kinematics.
With velocity--delay maps for several lines,
the radial ionization structure in the BLR becomes manifest,
and subtle structures such as spiral density waves may become
evident \citep{Horne04}.
Constructing velocity--delay maps was therefore the principal
motivation for undertaking the \storm\ campaign.

\subsection{ \memecho\ fits to the Spectral Variations}

Wavelength--delay maps $\Psi(\lambda,\tau)$ of the \new{emission-line response} 
in NGC~5548 are shown as two-dimensional false-color images 
in Fig.~\ref{fig:2dmdm} for the \memecho\ fit to the optical spectra from \mdm,
and in Fig.~\ref{fig:2dhst} for the UV spectra from \hst.
\new{ In the panel to the right of the 2D map,} the projections $\Psi(\tau)$
give delay maps for the full wavelength range (black) and for velocity ranges
centered on the rest wavelengths of the emission lines,
as indicated by the colored bars above and below the map. 
\new{ In the panel below, }  the projections $\Psi(\lambda)$ give the spectrum of the
full response (black) and of the response in four delay ranges,
0--5~days (purple), 5--10~days (green), 10--15~days~(orange), and 15--20~days (red). 
Velocity--delay maps centered on the six emission lines
are presented in Fig.~\ref{fig:2d}.
These two-dimensional maps show that the emission-line response
inhabits the interior of a virial envelope (dashed),
and exhibit structure indicating a Keplerian disk inclined by $i=45^\circ$ and
with an outer rim at $R/c=20$~days, as discussed below.
These maps and their interpretation are the main results of interest
emerging from our \memecho\ analysis.

\begin{figure*}
\begin{center}
\begin{tabular}{cc}
{\bf \hfill (a) \hfill ~} & \hfill {\bf (b) \hfill ~}
\\
 \includegraphics[angle=0,width=80mm]{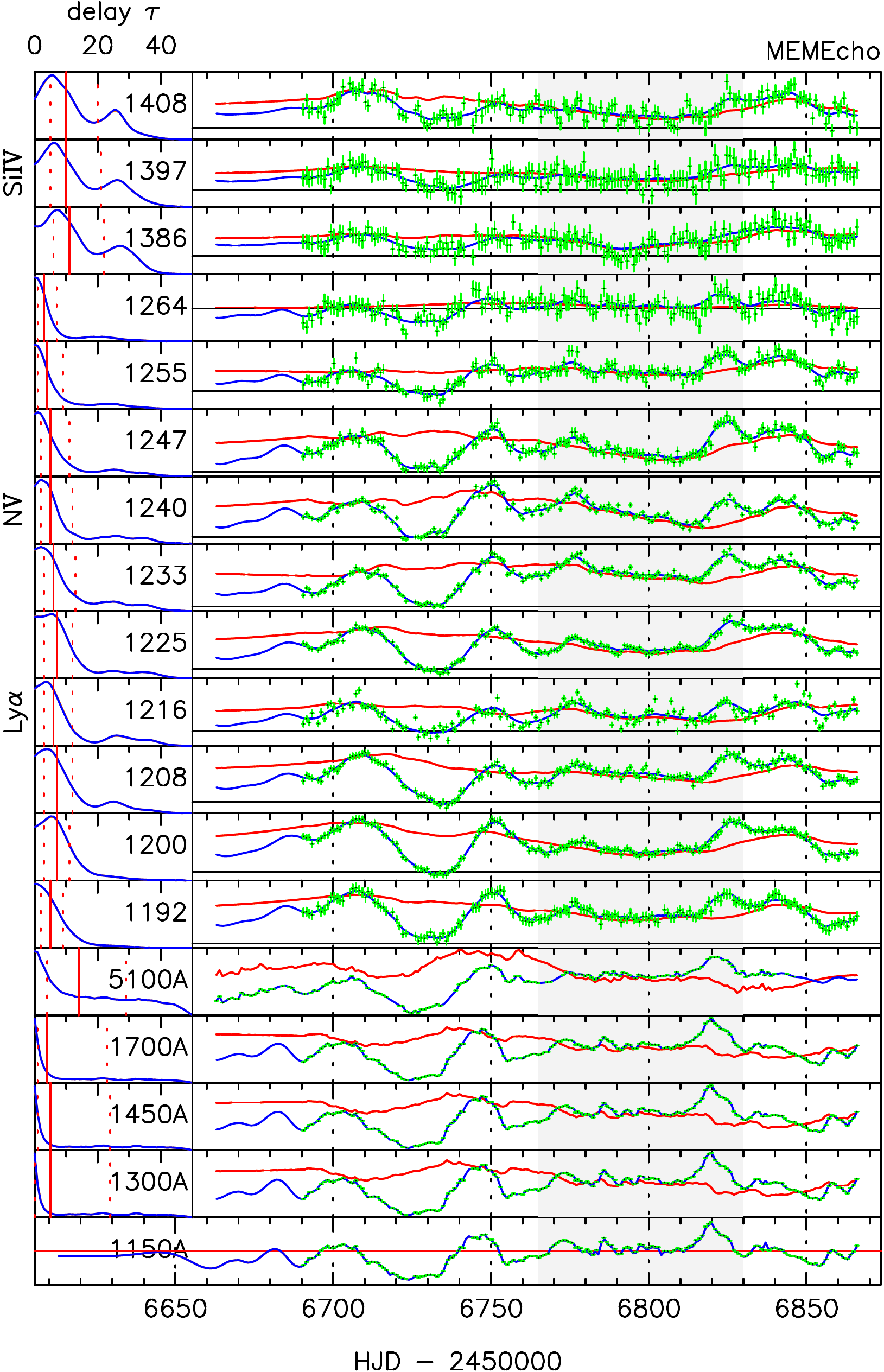}
&
 \includegraphics[angle=0,width=80mm]{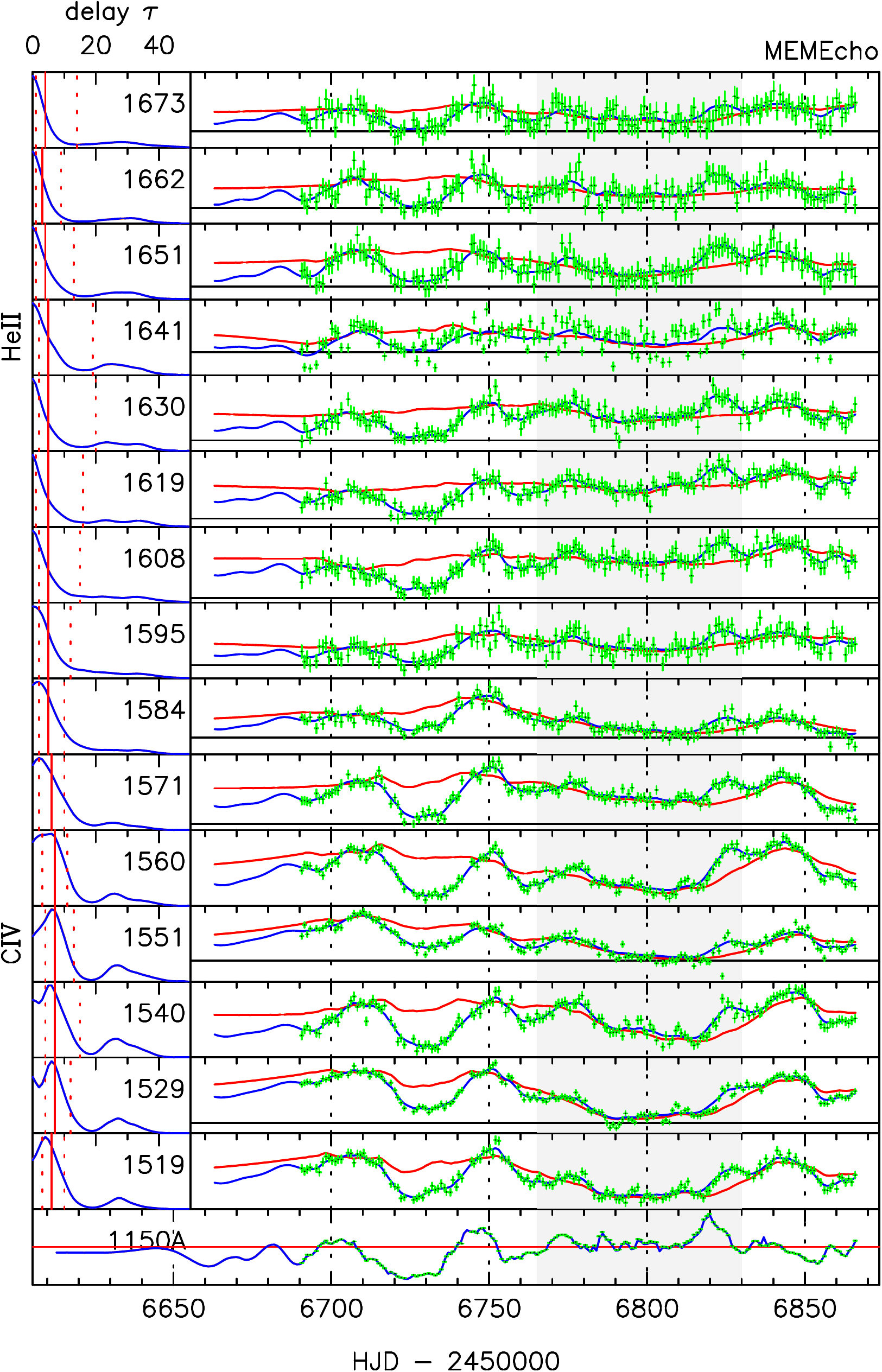}
\\
\end{tabular}
\end{center}
\caption{
Details of the \memecho\ fit to the spectral variations in the \hst\ data.
\new{Gray shading indicates the dates of the BLR Holiday.}
The light-curve data (black dots) with error bars (green) are compared with the
fitted model (blue) and varying background (red).
This global fit to $N=125,448$ data points achieved a reduced $\chi^2/N=1.2$.
The lower panel shows the driving light curve at 1150\,\AA.
Above this are delay maps (left) and echo light curves (right).
{\bf (a)} The continuum echos at 1300\,\AA, 1450\,\AA, 1700\,\AA, and 5100\,\AA,
and the continuum-subtracted emission at selected wavelengths,
including  \lya\,$\lambda1216$, \nv\,$\lambda1240$, and \siiv\,$\lambda1393$.
{\bf (b)} The echo maps and light curves 
for selected wavelengths including \civ\,$\lambda1549$ and \heii\,$\lambda1640$.
\label{fig:hstfit}
}
\end{figure*}

\begin{figure}
\begin{center}
  \includegraphics[angle=0,width=80mm]{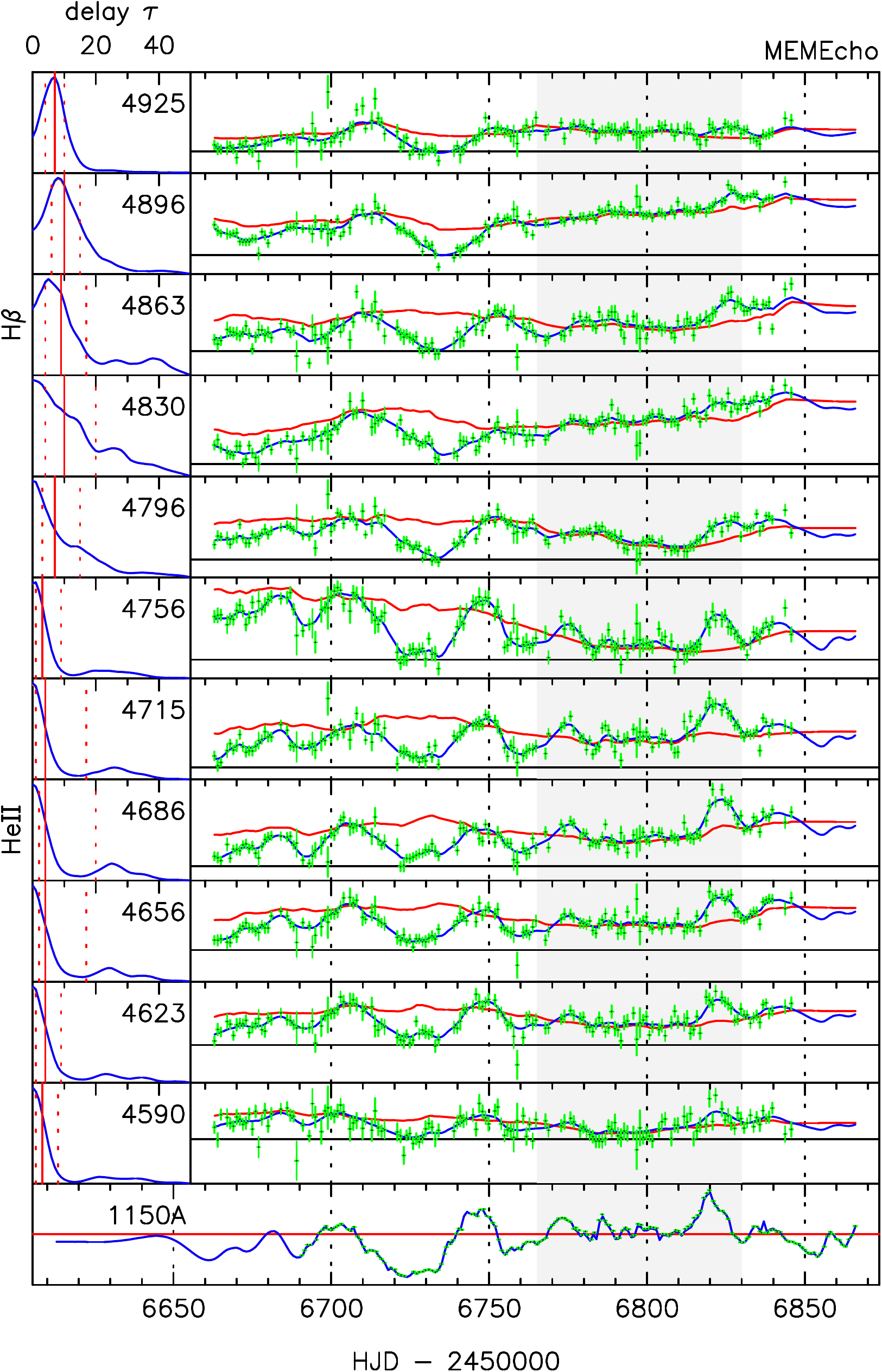}
\end{center}
\caption{
Details of the \memecho\ fit to the spectral variations in the \mdm\ data.
\new{Gray shading indicates the dates of the BLR Holiday.}
The light-curve data (black dots) with error bars (green) are compared with the
fitted model (blue) and the varying background (red).
This global fit to $N=125,448$ data points achieved a reduced $\chi^2/N=1.2$.
The lower panel shows the driving light curve at 1150\,\AA.
Above this are delay maps (left) and echo light curves (right)
for continuum echos at 1300\,\AA, 1450\,\AA, 1700\,\AA, and 5100\,\AA,
and for continuum-subtracted emission at selected wavelengths
including \heii\,$\lambda1640$, and \hb\,$\lambda4861$.
\label{fig:mdmfit}
}
\end{figure}

\begin{figure*}
\begin{center}
\begin{tabular}{cc}
{\bf \hfill (a) \hfill ~} & \hfill {\bf (b) \hfill ~}
\\
 \includegraphics[angle=270,width=0.49\textwidth]{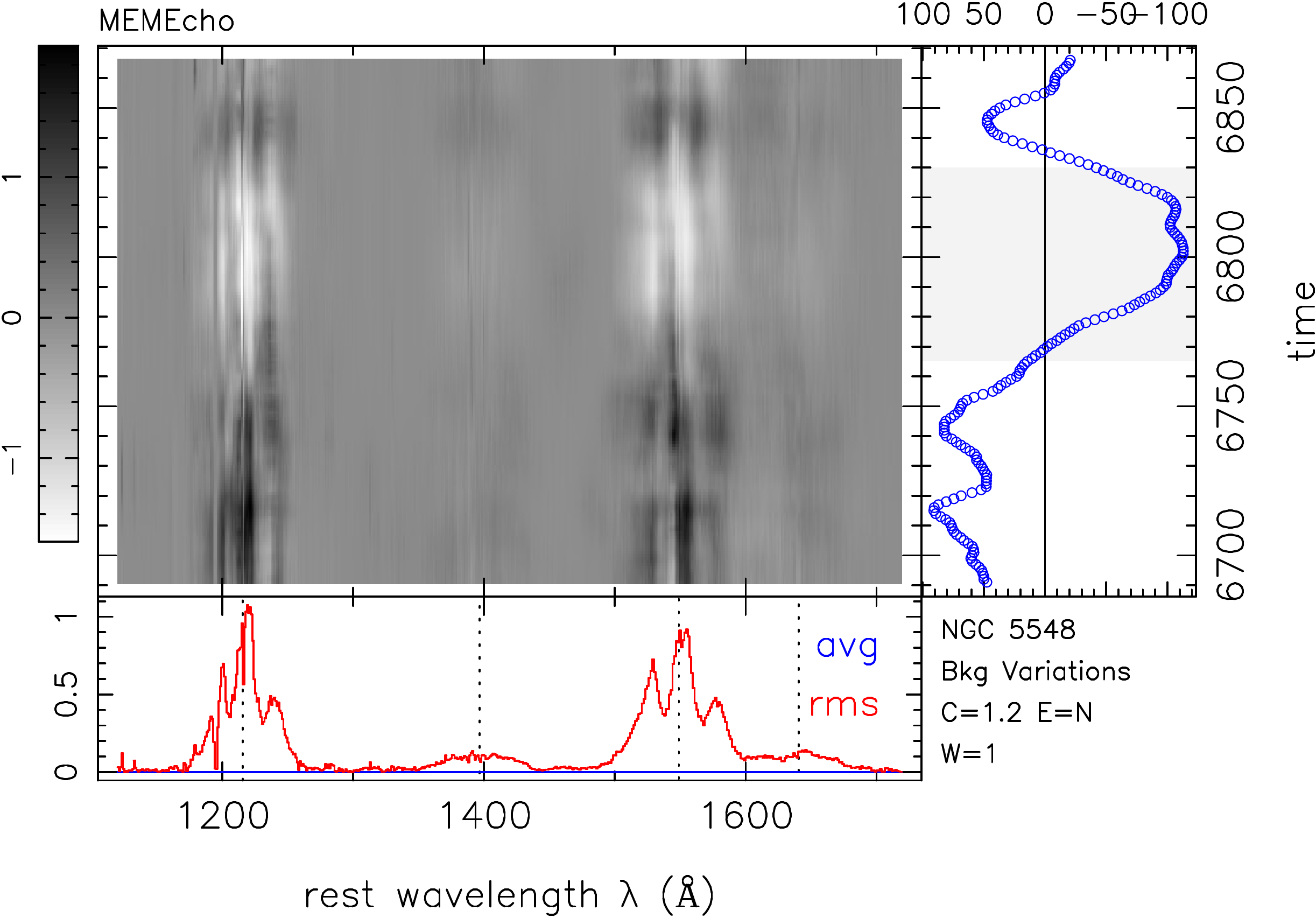}
&
 \includegraphics[angle=270,width=0.49\textwidth]{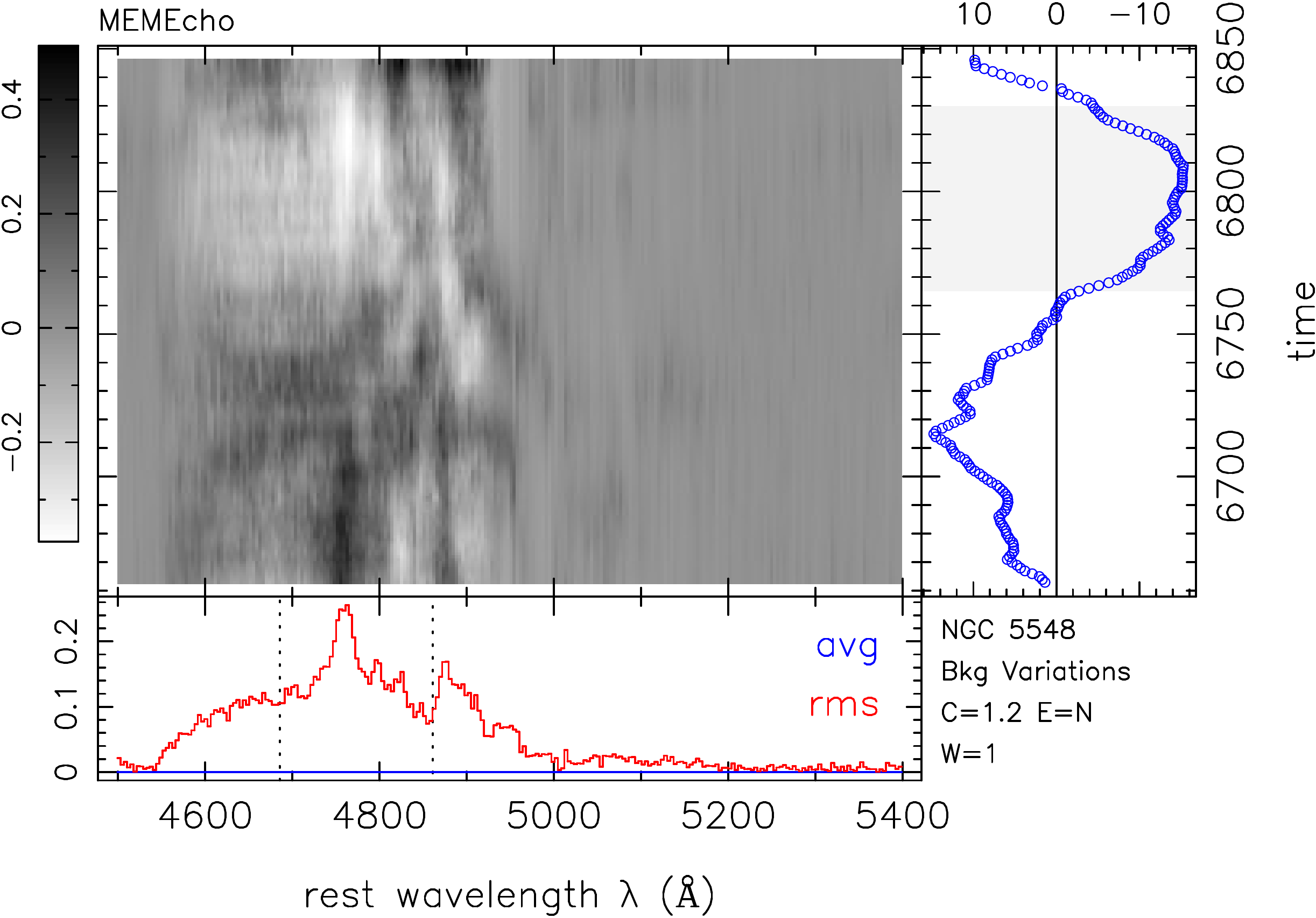}
\\
\end{tabular}
\end{center}
\caption{
The slowly-varying background component $L_0\left(\lambda,t\right)$ of the \memecho\
model fitted to the \hst\ data (left) and \mdm\ data (right).
The model's time-averaged spectrum has been subtracted
to leave time-varying residuals.
\new{The right panels show the wavelength-averaged response, with gray shading
indicating the BLR Holiday dates. 
The lower panels show the mean (blue) and rms (red) of the time-dependent residuals
at each wavelength.
All lines show a depressed flux during the BLR Holiday.}
The helical Barber-Pole pattern, with stripes moving from red to blue
across the line profile, is evident in the \civ\ and \lya\ residuals.
The Barber-Pole pattern is absent in \heii\, and may be present 
in \hb\ but with a less clear pattern.
\label{fig:slow}
}
\end{figure*}

Details from the \memecho\ fit are shown in
Figs~\ref{fig:hstfit} and \ref{fig:mdmfit} for the \hst\ and \mdm\ 
spectra, respectively, with the same
format as in Fig~\ref{fig:1d}.
The fit models as echoes
the same four continuum light curves
(at 1350\,\AA, 1450\,\AA, 1700~\AA, and 5100\,\AA), 
and now as well the continuum-subtracted emission-line variations
across the full wavelength range of the \hst\ and \mdm\ spectra. 
The \ps\ model provided the variable continuum model that was
subtracted to isolate the BLR spectra used in the \memecho\ fit.
As before, the proxy driving light curve $C(t)$ is the continuum
light curve at 1150\,\AA.
The model allows echo responses over a delay range 0--50~days,
and includes a time-variable background spectrum $L_0(\lambda,t)$.
This allows the model to account for the period of anomalous line 
response during the BLR Holiday (Paper~IV),
and any other features in the data, real or spurious, that are
not easily interpretable in terms of the linearized echo model.
Fig.~\ref{fig:slow} presents a grayscale ``trailed spectrogram''
display of the slowly-varying background $L_0(\lambda,t)$
for the \memecho\ fit to the \hst\ and \mdm\ spectra.
The Barber-Pole pattern of residuals and the BLR Holiday
features are evident here.

To regularize these fits, which include 2-dimensional wavelength-delay maps $\Psi(\lambda,\tau)$
and varying background spectra $L_0(\lambda,t)$,
the entropy now steers the fit toward models with smooth spectra
as well as smooth light curves and delay maps.
We actually construct a series of \memecho\ maps that fit
the data at different values of $\chi^2/N$, ranging from 5 to 1.
At higher $\chi^2$, the fit to the data is poor and the maps are smooth.
At lower $\chi^2$, the fit improves and the maps develop more detailed 
structure.
When $\chi^2$ is too low the fit becomes strained as the the model strives
to fit noise features (e.g., by introducing spikes in the gaps between data 
points in the driving light curve).
The fits and maps shown here for a fit with $\chi^2/N=1.2$ are
a good compromise between noise and resolution.
Our tests show that the main features interpreted here are robust to
changes in the control parameters of the fit. These parameters
adjust the relative ``stiffness'' of the driving light curve,
the background light curve, the echo maps, and the aspect 
ratio of resolution in the velocity and delay directions.

\subsection{Interpretation of the \memecho\ Maps}

 From Figs.~\ref{fig:2dmdm} and \ref{fig:2dhst},
the three strongest lines --- \lya, \civ, and \hb\ ---
have a similar velocity--delay structure,
with most of their response occurring between 5~days and 15~days.
To first order, the response in all three lines is red-blue symmetric. 
This indicates that radial motions are subdominant in the BLR, as
a strong inflow (outflow) component would produce shorter delays
on the red (blue) side of the velocity profile
\citep{Welsh91}.

 The \heii\ response is largely inside 5~days, and extends to $\pm10,000$~\kms,
compatible with expected radial ionization structure, and virial motions.
In Fig.~\ref{fig:2dmdm}, we see that \heii\ response is broad and
single-peaked. \heii\ dominates the \hb\ response in the 0--5-day
delay slice (purple), becomes subdominant at 5--10~days (green),
and almost negligible at larger delays.
The \heii\,$\lambda1640$ and \heii\,$\lambda4686$ delay ranges and velocity structures are
similar. There is no signature of a double-peaked structure in these two lines.

In Fig.~\ref{fig:2dmdm}, the \hb\ delay maps $\Psi(\tau)$ 
for the full profile ($\pm6000$~\kms, red) and for the line core
($\pm1500$~\kms, orange) are flat or rising from 0--10~days,
and then tail away.
The \hb\ response spectrum $\Psi(\lambda)$ exhibits
a double-peaked structure in the 10--15~day (orange)
and 15--20~day (red) delay ranges,
with the peaks separated by $\sim5000$~\kms.
In the 5--10-day slice, the \hb\ response has
a central peak flanked by ledges that extend to $\pm5000$~\kms.
Fig.~\ref{fig:2dhst} shows similar double-peaked
responses in \lya\ and somewhat less clearly in \civ.

Clearly recognizable in the velocity--delay structure is the
signature of an inclined Keplerian disk with a well-defined outer edge. 
The velocity--delay maps provide a plausible interpretation
for the ``M''-shaped variation in lag with velocity seen in the 
cross-correlation results (Papers I and V).
The outer edges of the ``M'' arise from the virial envelope.
The ``U''-shaped interior of the ``M'' dips down from 20~days to 5~days,
and we interpret this as the lower half of an
ellipse in the velocity--delay plane, which is the signature
of a ring of gas orbiting the black hole at radius $R=20$~light days.

 The \hb\ response exhibits the clearest signature of an ellipse
in the velocity--delay plane,
corresponding to an annulus in the Keplerian disk.
The (stronger) near side of the annulus has a delay at 
$\tau=(R/c)(1-\sin{i})\approx5$~days and the (weaker) far side extends to 
$\tau=(R/c)(1+\sin{i})\approx35$~days
Assuming a thin disk, the ratio gives $\sin{i}\approx 0.75$, or $i\approx45^\circ$.
The double-peaked velocity structure at $\tau\approx20$\,days then gives the black hole mass. 
The framework shown by dashed orange curves on Figs~\ref{fig:hstfit} and \ref{fig:mdmfit}
were adjusted by eye to fit the main features.
This provides rough estimates for the black hole mass $\mbh\approx7\times10^7~\msun$,
the inclination $i\approx45^\circ$, and the outer BLR radius $R_{\rm out}\approx20$~light days.
Note that the characteristic BLR response timescale,
as measured by the mode or mean or median of $\Psi(\tau)$, 
is less than $R/c$ at the outer edge of the BLR.

The velocity--delay structure also indicates a stronger response
from the near side than from the far side of the inclined disk.
We see the upper half of the ellipse only faintly in the velocity--delay
map of \hb\ and perhaps also for \lya.
The \civ\ map and the (less reliable) \siiv\ map show a faint response at 25--30~days
that is not very clearly connected to the stronger response inside 10--15~days.
If the response structure were azimuthally symmetric, the upper and lower
halves of the ellipse would be more equally visible.
The mean delay averaged around the ellipse would be $R/c\approx20$~days,
and this is similar to typical \hb\ lags seen in the past.
The much shorter lags in the \storm\ data may be interpreted
as due at least in part to an anisotropy present in 2014
that was usually much weaker or absent
during previous monitoring campaigns.
The near/far contrast ratio can be determined by more careful modeling.

\section{Discussion}
\label{sec:previous}

\subsection{ Parameter Uncertainties}

The morphology of the velocity--delay maps indicates that
the BLR in \ngc\ is compatible with a disk-like geometry and kinematics,
for a black hole mass $\mbh\approx7\times10^7\,\msun$,
an inclination angle $i\approx45^\circ$, and with a relatively
sharp outer rim at $R/c\approx20$~d.
Uncertainties in these parameters are difficult to quantify precisely
because \memecho\ delivers  the ``smoothest positive'' maps that
fit the data at specified levels of $\chi^2/N$.
This relatively model-independent approach  does not aim
to determine model-specific parameters.
However, by comparing the velocity--delay maps with those of toy models, 
we estimate that \mbh\  is uncertain by $\sim20$\%,
the inclination by $\sim 10^\circ$, and $R/c$ by $\sim10$\%.

Future and ongoing efforts to model the \storm\ data will incorporate
more specific geometric and kinematic parameters
and photoionization physics.
These should allow for parameter estimates with 
better-quantified uncertainties.
At this stage the value of the velocity--delay maps is to 
inform the dynamical modeling efforts by indicating
what types of models are likely to succeed.

\subsection{ Comparison with 2008 mass estimates }

\ngc\ was one of 13 AGN monitored during the 
the Lick AGN Monitoring Project (\lamp) 2008 
reverberation mapping campaign \citep{Bentz09lamp,Walsh09}.
Using spectra from the 3\,m Shane telescope
at Lick Observatory and Johnson $V$ and $B$ broad-band photometry
from a number of ground-based telescopes,
the \lamp~2008 campaign
secured 57 $V$-band continuum and 51 \hb\ line epochs on \ngc\
over a 70-day span.
The centroid \hb\ lag measured by cross-correlation methods
is $\tau_{\rm cent}=4.2^{+0.9}_{-1.3}$~days.
Combined with the \hb\ linewidth
 $\sigma_{\rm line}=4270\pm290$~km~s$^{-1}$ from the rms spectrum,
 the virial product is
 $c\, \tau_{\rm cent}\, \sigma_{\rm line}^2/G=14.9^{+3.7}_{-5.1}\times10^6$~\msun,
 giving the black hole mass as
 $\mbh=8.2^{+2.0}_{-2.8}\times10^7\,\left(f/5.5\right)$~\msun,
 where $f$ is the adopted calibration factor, uncertain by $\sim0.3$ dex
 for individual AGN.

\cite{Pancoast14} present results of a dynamical modeling fit to
the 2008 \lamp\ data on NGC~5548. 
The \caramel\ code employs Markov Chain Monte-Carlo methods
to sample the parameters of a dynamical model of the \hb-emitting region.
Taking the $V$-band data as a proxy for the driving light curve, \caramel\
fits the reverberating emission-line profile variations by adjusting
the spatial and kinematic distribution of \hb-emitting clouds.
This model incorporates more detailed model-specific information
than the \memecho\ mapping, but is flexible enough to explore
a variety of inflow and outflow as well as disk-like kinematic structures.
The \caramel\ fit to  \ngc\ in 2008 finds a mean delay
of just 3~days, indicating a smaller \hb\ emission-line region.
 The \caramel\ model fitted to the 2008 data has a thick-disk geometry,
 with an opening angle $\theta_0=(27^{+11}_{-8})^\circ$,
 a dominant \hb\ response on the far side of the disk,
 and a strong inflow signature, with small
 delays on the red wing relative to those on the blue wing of \hb.
 The inflow and far-side response in 2008 are quite different from the symmetric disk-like kinematics 
 and near-side response evident in the \memecho\ velocity--delay maps.
Despite these differences, and 
the smaller size of the \hb\ emission-line region in 2008,
the inferred mass $\mbh=3.9^{+2.9}_{-1.5}\times10^7$~\msun\
and inclination  $i=(39\pm12)^\circ$ are consistent,
given their uncertainties, with the corresponding values,
$\mbh\approx7\times10^7$~\msun\ and $i\approx45^\circ$,
that we estimate from the structure of the \hb\
velocity--delay maps from the  2014 \storm\ data.
The mass estimate from the \caramel\ fit may be too small
due to the \hb\ response being measured relative to the optical
continuum, which itself may have a delay of a few days.
For example, in 2014 the $V$-band delay was $1.6\pm0.5$~days (Paper~II),
and that could boost the \caramel\ mass estimate by $\sim50$\%,
bringing it into closer agreement with our estimate from the
velocity--delay maps.

\subsection{ Comparison with 2015 velocity--delay maps}

During 2015~Jan-Jul, a year after the \storm\ campaign, \ngc\ was monitored with the 
Yunnan Faint Object Spectrograph and Camera on the 2.4~m
telescope at Lijiang, China, resulting in 61 good spectra over 205 days 
that provide the basis for a MEM analysis yielding a velocity--delay map for \hb\
\citep{Xiao18}. This 2015 map exhibits structure remarkably similar 
to that seen in our 2014 map, including the virial envelope,
the ``M''-shaped structure with 
$\tau\approx10$~d and $V\approx\pm2400$~\kms\ at the peaks of the ``M,''
and a well-defined ellipse extending out to $\tau\approx40$~d.

\cite{Xiao18} also present \hb\ velocity--delay maps constructed
from 13 annual AGNWatch campaigns 1989--2001.
While several of these maps show hints of a ring-like structure,
the quality of these maps is much lower owing to less intensive
time sampling, making it difficult to be confident about 
the information that they may be able to convey.

In combination, the high-fidelity 2014 and 2015 \hb\ velocity--delay maps
each show a clear virial envelope and distinct ellipse centered at 20 light days.
Thus, at both epochs the \hb\ response arises from a Keplerian disk
with a relatively sharp outer rim at $20$ light days that remained
stable for an interval of at least a year.
In 2014 the response is weak on the top of the velocity--delay ellipse.
In 2015 the response is clearly visible on the blue side
and over the top of the velocity--delay ellipse, and relatively weak
on the red side.
This indicates significant azimuthal modulation of the response
that evolves, perhaps rotates, on a timescale of a year.
Future velocity--delay mapping experiments to monitor this
structure could be interesting to elucidate its origin and implications.

\subsection{ Barber-Pole patterns}

The Barber-Pole pattern uncovered here in \ngc\ may be related 
to intermittent periodic phenomena seen in the
subclass of AGN that have double-peaked Balmer emission-lines.
In these double-peaked emitters there is a clear separation between the narrow emission lines
and the broad double-peaked lines.
The double-peaked velocity profiles can be modeled by emission
from a Keplerian disk,  with relativistic effects making the blue peak stronger 
and sharper than the red one and redshifting the line center.
Spectroscopic monitoring of Arp~102B during 1987--1996 detected \hb\ profile variations
\citep{Newman97}.
In particular, during 1991--1995, the \hb\ red/blue flux ratio oscillates over 
nearly 2 cycles of a 2.2~yr period,
suggesting a patch of enhanced emission on a circular orbit within the disk.
Contemporaneous monitoring  of \ha\ during 1992-96,
shows that its red/blue flux ratio also oscillates
by $\pm10$\% with a 2~yr period \citet{Sergeev00}.
A trailed spectrogram display of the \ha\ residuals, 
after subtracting scaled mean line profiles, 
reveals a helical Barber-Pole pattern  with 2 stripes that move from red to blue
across the double-peaked \ha\ profile.
This is strikingly similar to what we see in the \civ\ profile of \ngc.
Similar phenomena are seen in other double-peaked emitters
\citep{Gezari07, Lewis10, Schimoia15, Schimoia17}.
Our discovery of the Barber-Pole pattern in \ngc\ 
indicates that this phenomenon is not limited
to the double-peaked emitters.

\section{Summary}

\label{sec:fini}

In this paper, we achieve the primary goal of the AGN \storm\ campaign
by recovering velocity--delay maps
for the prominent, broad, \lya, \civ, \heii, and \hb\ emission lines
in  \ngc.
These are the most detailed velocity--delay maps yet obtained for an AGN,
providing unprecedented information on
the geometry, ionization structure, and kinematics of the broad-line region.

Our analysis interprets the ultraviolet \hst\ spectra
(Paper~I) and optical \mdm\ spectra (Paper~V) secured
in 2014 during the 6-month \storm\ campaign on \ngc.
This dataset provides spectrophotometric
monitoring of \ngc\ with unprecedented duration, cadence,
and S/N suitable for interpretation
in terms of reverberations in the broad emission-line
regions (BLR) surrounding the black hole.
Assuming that the time delays arise from light travel time,
the velocity--delay maps we construct from the
reverberating spectra provide 2-dimensional projected images
of the BLR, one for each line, resolved on
iso-delay paraboloids and line-of-sight velocity.

We used the absorption line modeling results from Paper~VIII
to divide out absorption lines affecting the \hst\ spectra.
We used \ps\ to recalibrate the flux, wavelength, and spectral
resolution of the optical \mdm\ spectra using the strong narrow 
emission lines as internal calibrators.
Residuals from the \ps\ fits indicate the success of the calibration
adjustments.

The linearized echo model that we normally use for echo mapping
is violated in the \storm\ dataset by anomalous emission-line behavior,
the BLR Holiday discussed in Paper~IV.
We model this adequately as a slowly varying background spectrum
superimposed on which are the more rapid variations due to
reverberations.

The residuals of the \ps\ fits reveal significant
emission-line profile changes. Features are evident
moving inward from both red and blue wings
toward the center of the \hb\ line, interpretable as
reverberations of a BLR with a Keplerian velocity field.

A helical ``Barber-Pole'' pattern with stripes moving from
red to blue across line profile is evident in the \civ\ and 
possibly also the \lya\ lines, suggesting an azimuthal
structure rotating with a period of $\sim2$~yr around the
far side of the accretion disk.
This may be due to precession or orbital
motion of disk structures.
Similar behavior is seen in the double-peaked emitters, such as Arp~102B.
Further \hst\ observations of \ngc\ over a multi-year timespan,
with a cadence of perhaps 10~days rather than 1~day,
could be an efficient way to explore the persistence, transience
and nature of this new phenomenon in \ngc.

We use the \ps\ fit to extract light curves for the lines and continua
and use \memecho\ to fit these light curves
using the 1150\,\AA\ continuum light curve as a proxy
for the driving light curve.
The \memecho\ fit determines a set of echo maps $\Psi(\tau)$
giving the delay distribution of each echo, effectively
slicing up the reverberating region on iso-delay paraboloids.
The structure in these echo maps indicates radial stratification,
with \heii\ responding from inside 5~light days, and the
\lya\, \civ, and \hb\ response extending out to 
or beyond 20~light days.

By using \memecho\ to fit reverberations in the emission-line profiles,
we construct velocity--delay maps $\Psi(v,\tau)$ that resolve
the BLR in time delay and line-of-sight velocity.
The BLR response is confined within a virial envelope around
each line, with double-peaked velocity profiles in the
response in 10--20 day delay slices. The ``M"-shaped 
changes in delay with velocity, found using 
velocity-resolved cross-correlation lags
in Papers~I and V, are seen here to be the 
signature of a Keplerian disk. The outer legs of the ``M''
arise from the virial envelope between 5 and 20 days,
the inner ``U'' of the ``M'' is the lower part of an ellipse
extending from 5 to 35 days.
This velocity--delay structure is most straightforwardly
interpreted as arising from a Keplerian disk
extending from $R/c=5$ to 20 days, inclined by
$i\approx45^\circ$, and centered on a black hole 
of mass $\mbh\approx7\times10^7~\msun$.
The BLR has a well-defined outer rim at $R/c\approx20$ days,
but the far side of the rim may be obscured or less
responsive than the near side.

Detailed modeling of the \storm\ data, guided by
the features in the velocity--delay maps presented here,
should be able to refine and quantify uncertainties on
these features of the BLR, the inclination, and the black hole
mass. 

\vspace*{3mm}

{\bf ACKNOWLEDGEMENTS}

\acknowledgments
Support for \hst\ program number GO-13330 was provided by NASA through a
grant from the Space Telescope Science Institute, which is operated by the
Association of Universities for Research in Astronomy, Inc., under NASA
contract NAS5-26555.
~
K.H.\ acknowledges support from STFC grant ST/R000824/1.
~
G.D.R., B.M.P., M.M.F, C.J.G., and R.W.P.\ are grateful for the support of the
National Science Foundation (NSF) through grant AST-1008882 to
The Ohio State University.
~
Research at UC~Irvine has been supported by NSF grants AST-1412693 and AST-1907290.
~
M.C.Bentz gratefully acknowledges support through NSF CAREER grant AST-1253702
to Georgia State University.
~
V.N.B.\ gratefully acknowledges assistance from National Science Foundation
(NSF) Research at Undergraduate Institutions (RUI) grant AST-1909297.
~
S.B.\ is supported by NASA through Chandra award AR7-18013X issued by
the Chandra X-ray Observatory Center, operated by the Smithsonian
Astrophysical Observatory for and on behalf of NASA under contract
NAS8-03060. S.B.\ was also partially supported by grant HST-AR-13240.009.
~
M.C.\ Bottorff acknowledges HHMI for support through an undergraduate science
education grant to Southwestern University.
~
E.M.C., E.D.B., L.M., and A.P.\ acknowledge support from Padua University 
through grants DOR1699945/16, DOR1715817/17, DOR1885254/18, and BIRD164402/16.
~
G.F.\ and M.\ Dehghanian acknowledge support from the NSF (AST-1816537), NASA (ATP 17-0141),
and STScI (HST-AR-13914, HST-AR-15018), and the Huffaker Scholarship.
~
K.D.D.\ is supported by an NSF Fellowship awarded under grant AST-1302093.
R.E.\ gratefully acknowledges support from NASA under ADAP award
80NSSC17K0126.
~
P.A.E.\ acknowldges UKSA support.
~
Support for A.V.F.'s group at U.C.Berkeley is provided by
NSF grant AST-1211916,
the TABASGO Foundation, 
the Christopher R. Redlich Fund, 
and the Miller Institute for Basic Research in Science.
~
J.M.G.\ gratefully acknowledges support from NASA under awards
NNX15AH49G and 80NSSC17K0126.
~
P.B.H.\ is supported by NSERC.
~
M.I.\ acknowledges support from the National Research Foundation of Korea (NRF) grant, No.\
2020R1A2C3011091.
~
M.D.J.\ acknowledges NSF grant AST-0618209.
~
M.K.\ was supported by the National Research Foundation of Korea (NRF) grant
funded by the Korea government (MSIT) (No.\ 2017R1C1B2002879).
SRON is financially supported by NWO, the
Netherlands Organization for Scientific Research.
~
C.S.K.\  is supported by NSF grants 
AST-1515876 and AST-1814440.
~
Y.K.\ acknowledges support from  DGAPA-PAIIPIT grant IN106518.
~
D.C.L.\ acknowledges support from NSF grants AST-1009571 and AST-1210311.
~
P.L.\ acknowledges support from Fondecyt grant 1120328.
~
A.P.\ is supported by NASA through Einstein Postdoctoral Fellowship grant number PF5-160141
awarded by the Chandra X-ray Center, which is operated by the Smithsonian Astrophysical Observatory
for NASA under contract NAS8-03060.
~
J.S.S.\ acknowledges CNPq, National Council for Scientific and Technological
Development (Brazil) for partial support and The Ohio State University
for warm hospitality.
~
T.T.\ has been supported by NSF grant AST-1412315.
T.T.\ and B.C.K.\ acknowledge support from the Packard Foundation in the form of
a Packard Research Fellowship to T.T.
The American Academy in Rome and the
Observatory of Monteporzio Catone are thanked by T.T.\ for kind hospitality.
~
\new{MV gratefully acknowledges support from the Independent Research Fund Denmark 
via grant numbers DFF 4002-00275 and 8021-00130. }
~
J.-H.W.\ acknowledges support by the National Research Foundation of Korea (NRF)
grant funded by the Korean government (No.\ 2010-0027910).
~
This research has made use of the NASA/IPAC Extragalactic Database (NED),
which is operated by the
Jet Propulsion Laboratory, California Institute of Technology,
under contract with the National Aeronautics and Space Administration.
Research at Lick
Observatory is partially supported by a generous gift from Google.


\begin{thebibliography}{}
\expandafter\ifx\csname natexlab\endcsname\relax\def\natexlab#1{#1}\fi





\bibitem[Bentz \et\ (2009)]{Bentz09lamp}
Bentz, M.C. \et\ 2009, \apj, 705, 199.
 







\bibitem[{{Blandford} \& {McKee}(1982)}]{Blandford82}
{Blandford}, R.~D., \& {McKee}, C.~F. 1982, \apj, 255, 419




\bibitem[Cackett, Horne, \& Winkler(2007)]{Cackett07}
Cackett, E.M., Horne, K., \& Winkler, H. 2007, \mnras, 380, 669



\bibitem[{{Chelouche} \et\ (2019)}]{Chelouche19}
Chelouche D., Pozo Nunez F., Kaspi S., 2019, Nat.~Astron., 3, 251





\bibitem[Collier \et\ (2001)]{Collier01}
Collier, S.J., \et\ 2001, \apj, 561, 146






\bibitem[Dehghanian \et\ (2019a)]{Dehghanian19}
Dehghanian, M., \et\ 2019, \apj\, 877, 119 (Paper~X)

\bibitem[Dehghanian \et\ (2019b)]{Dehghanian19b}
Dehghanian, M. \et\ 2019 \apjl, 882, L30

\bibitem[De Rosa \et~(2015)]{DeRosa15}
De Rosa, G., Peterson, B.M., Ely, J., et~al. 2015, \apj, 806, 128 (Paper~I)

	





\bibitem[Edelson \et\ (2015)]{Edelson15}
Edelson, R., Gelbord, J.M., Horne, K., \et~ 2015, \apj, 810, 129 (Paper~II)



\bibitem[Eracleous \& Halpern(1994)]{Eracleous94}
Eracleous, M., \& Halpern, J.P. 1994, \apjs, 90, 1

\bibitem[Eracleous \& Halpern(2003)]{Eracleous03}
Eracleous, M., \& Halpern, J.P. 2003, \apj, 599, 886


\bibitem[Fausnaugh \et\ (2016)]{Fausnaugh16}
Fausnaugh, M.M., Denney, K.D., Barth, A.J., et~al. 2016, \apj, 821, 56 (Paper~III)

\bibitem[Gardner \& Done(2017)]{Gardner17}
Gardner, E., \& Done, C. 2017, \mnras, 470, 2245






\bibitem[Gezari \et\ (2007)]{Gezari07}
Gezari, S., Halpern, J.P., \& Eracleous, M. 2007, \apjs, 169, 167

\bibitem[Goad \et\ (2016)]{Goad16}
Goad, M.R., Korista, K.T., De Rosa, G., et~al., \apj, 824, 11 (Paper~IV)






\bibitem[Grier \et\ (2013)]{Grier13}
Grier, C.J., Peterson, B.M., Horne, K., \et\ 2013, \apj, 764, 47







\bibitem[{{Horne}(1994)}]{Horne94}
{Horne}, K. 1994, in Astronomical Society of the Pacific Conference Series,
  Vol.~69, Reverberation Mapping of the Broad-Line Region in Active Galactic
  Nuclei, ed. P.~M. {Gondhalekar}, K.~{Horne}, \& B.~M. {Peterson}, pp.\ 23--51


\bibitem[Horne \et\ (2004)]{Horne04}
Horne, K., Peterson, B.M., Collier, S.J., \& Netzer, H., 2004, 
\pasp, 116, 465


\bibitem[Jarvis \& McLure(2006)]{Jarvis06}
Jarvis, M.J., \& McLure, R.J. 2006, \mnras, 369, 182











\bibitem[Korista \& Goad (2019)]{Korista19}
Korista, K.T., \& Goad, M.R. 2019, \mnras, 489, 5284




\bibitem[Kriss \et\ (2019)]{Kriss19}
Kriss, G.A., De~Rosa, G., Ely, J., Peterson, B.M.,
\et\ 2019, \apj, 881, 153 (Paper~VIII)




\bibitem[Lawther \et\ (2018)]{Lawther18}
Lawther D., Goad M.R., Korista K.T., Ulrich O., Vestergaard M., 2018, \mnras, 481, 533

\bibitem[Lewis, Eracleous, \& Storchi-Bergmann(2010)]{Lewis10}
Lewis, K.T., Eracleous, M., \& Storchi-Bergmann, T. 2010, \apjs, 187, 416




\bibitem[Mangham \et\ (2019)]{Mangham19}
Mangham, S.W., \et\ 2019, \mnras, 488, 2780. 

\bibitem[Manser \et\ (2019)]{Manser19}
Manser, C.M., \et\ 2019, Science, 364, 66.


\bibitem[Marconi \et\ (2008)]{Marconi08}
Marconi, A., Axon, D.J., Maiolino, R., Nagao, T., Pastorini, G.,
Pietrini, P., Robinson, A., \& Torricelli, G. 2008, \apj, 678, 693



\bibitem[Mathur \et\ (2017)]{Mathur17}
Mathur, S., Gupta, A., Page, K., Pogge, R.W., Krongold, Y., Goad, M.R.,
\et\ 2017, \apj, 846, 55 (Paper~VII)





\bibitem[Morgan \et\ (2010)]{Morgan10}
Morgan C.W., Kochanek C.S., Morgan N.D., Falco E.E., 2010,  \apj, 712, 1129

\bibitem[Mosquera \et\ (2013)]{Mosquera13}
Mosquera A.M., Kochanek C.S., Chen B., Dai X., Blackburne J.A., 
Chartas G., 2013, \apj, 769, 53

\bibitem[Mummery \& Balbus (2020)]{Mummery20}
Mummery, A. \& Balbus, S.A. 2020, \mnras, 492, 5655




\bibitem[Netzer \& Marziani (2010)]{Netzer10}
Netzer, H., \& Marziani, P.,
2010, \apj, 724, 318


\bibitem[Newman \et\ (1997)]{Newman97}
Newman, J.A., Eracleous, M., Filippenko, A.V., Halpern, J.P., 1997, \apj\ 485, 570



\bibitem[Pancoast \et\ (2014)]{Pancoast14}
Pancoast, A., Brewer, B.J., Treu, T., Park, D., Barth, A.J.,
Bentz, M.C., \& Woo, J.-H. 2014, \mnras, 445, 3073








\bibitem[Pei \et\ (2017)]{Pei17}
Pei, L., Fausnaugh, M.M., Barth, A.J., et~al. 2017, 
\apj, 837, 131 (Paper~V)

\bibitem[{{Peterson}(1993)}]{Peterson93}
{Peterson}, B.M. 1993, \pasp, 105, 247


\bibitem[{{Peterson}(2014)}]{Peterson14}
Peterson, B.M. 2014, Space Sci.\ Rev., 183, 253







\bibitem[Peterson et~al.(2004)]{Peterson04}
Peterson, B.M., {et~al.}  2004, \apj, 613, 682


\bibitem[Peterson \et\ (2013)]{Peterson13}
Peterson, B.M., \et\  2013, \apj, 779, 109


\bibitem[Poindexter \et\ (2008)]{Poindexter08}
Poindexter, S., Morgan, N., Kochanek, C.S., 2008, \apj, 673, 34







\bibitem[Schimoia \et\ (2015)]{Schimoia15}
Schimoia, J.S., Storchi-Bergmann, T., Grupe, D., Eracleous, M.,
Peterson, B.M., Baldwin, J.A., Nemmen, R.S., 
Winge, C., 2015, \apj, 800, 63

\bibitem[Schimoia \et\ (2017)]{Schimoia17}
Schimoia, J.S., Storchi-Bergmann, T., 
Winge, C., Nemmen, R.S.,  Eracleous, M., 
2017 \mnras, 472, 2170
 


\bibitem[Sergeev \et\ (2000)]{Sergeev00}
 Sergeev, S.G., Pronik, V.I., Sergeeva, E.A., 2000, A\&A, 356, 41
 





\bibitem[Shakura \& Sunyaev (1973)]{SS73}
Shakura, N.I., \& Sunyaev, R.A. 1973, A\&A, 24, 337




\bibitem[Smith \et\ (2004)]{Smith04}
Smith, J.D., \et\ 2004, \mnras, 350, 140

\bibitem[Starkey \et\ (2016)]{Starkey16}
Starkey, D., Horne, K., Fausnaugh, M.M., 
\et~2016, \apj, 835, 65 (Paper~VI)

\bibitem[Storchi-Bergmann \et\ (2017)]{Storchi-Bergmann17}
Storchi-Bergmann, T., Schimoia, J.S., Peterson, B.M., Elvis, M., Denney, K.D.,
Eracleous, M., \& Nemmen, R.S., 2017, \apj, 835, 236

\bibitem[Sturm \et\ (2018)]{Sturm18}
Sturm, E., \et\ 2018, Nature, 564, 657

\bibitem[Strateva \et\ (2003)]{Strateva03}
Strateva, I.V., \et\ 2003, \aj, 126, 1720


\bibitem[Sun \et\ (2020)]{Sun20}
Sun, M., \et\ 2020, arXiv:2002.08564








\bibitem[Vestergaard, Wilkes, \& Barthel~(2000)]{Vestergaard00}
Vestergaard, M., Wilkes, B.J., \& Barthel, P.D. 2000, \apj, 302, 56


\bibitem[Walsh \et\ (2009)]{Walsh09}
Walsh, J.L., \et\ 2009, \apjs, 185, 156





\bibitem[{Welsh} \& {Horne} (1991)]{Welsh91}
Welsh, W.F., \& Horne, K. 1991, \apj, 379, 586




\bibitem[Wills \& Browne(1986)]{Wills86}
Wills, B.J., \& Browne, I.W.A. 1986, \apj, 302, 56


\bibitem[Xiao \et\ (2018)]{Xiao18}
{Xiao}, M., \et\ 2018, \apjl, 865, L8

\bibitem[Young \et\ (2007)]{Young07}
Young, S., \et\ 2007, Nature, 450, 74






\end{thebibliography}
\end{document}